\documentclass[journal]{IEEEtran}
\usepackage[cmex10]{amsmath}
\usepackage{cite}
\usepackage{color}
\usepackage{placeins}
\usepackage{float}
\usepackage{tabularx,colortbl}
\usepackage{graphicx}
\usepackage{epstopdf}
\usepackage{subfigure}	
\usepackage{tabularx,colortbl}
\usepackage{comment}
\usepackage{stfloats}
\usepackage{esint}
\usepackage{xcolor}
\usepackage{makecell}
\def\plotsize{0.34}
\usepackage{hyperref}

\begin{document}

\title{\textcolor{black}{Toward a Distributed Radio Telescope Using Global IoT Networks: Calibration Methods and Feasibility Analysis}}

\author{Junming Diao,~\IEEEmembership{Senior Member,~IEEE}
\thanks{The authors Junming Diao is with the Department of Electrical and Computer Engineering, Mississippi State University, Starkville, MS, USA, E-mail: jdiao@ece.msstate.edu.}
}
\maketitle


\begin{abstract}
\textcolor{black}{This paper presents a conceptual feasibility study of leveraging large-scale Internet of Things (IoT) networks to form a distributed radio telescope for astronomical observations.} Leveraging existing IoT infrastructure with minimal modifications, the proposed receiving system employs widely dispersed devices to capture both astronomical and communication signals at the same time and frequency. A novel calibration approach is proposed using multiple distributed satellites transmitting known signals, enabling precise channel estimation and phase correction via GPS localization. A system signal model is employed to study different calibration algorithms with node hardware variations, node clock offsets, and interference mitigation. Compared to the 500-meter Aperture Spherical Telescope (FAST), \textcolor{black}{assuming the ideal calibration and sufficient number of node,} the IoT-enabled telescope increases survey speed by seven orders of magnitude, owing to the large antenna gain and expansive field of view (FoV). These findings demonstrate the feasibility of future radio telescope architectures that leverage large-scale distributed nodes, and highlight their potential advantages in terms of low cost, high survey speed performance, and spectrum-sharing capability.

\end{abstract}

\IEEEpeerreviewmaketitle

\section{Introduction}
\IEEEPARstart{R}{A}DIO astronomy is a cornerstone of modern astrophysics, harnessing radio frequency observations to probe celestial phenomena including stars, galaxies, quasars, and pulsars. At the heart of this discipline are radio telescopes, which typically employ single reflector antenna or multi-reflector arrays to collect extremely faint astronomical signals. These signals are then directed to a feed-receiving system located at the reflector's focal point, where they undergo subsequent signal processing and analysis.

\begin{figure}[htp]
\centering
\includegraphics[scale=0.25]{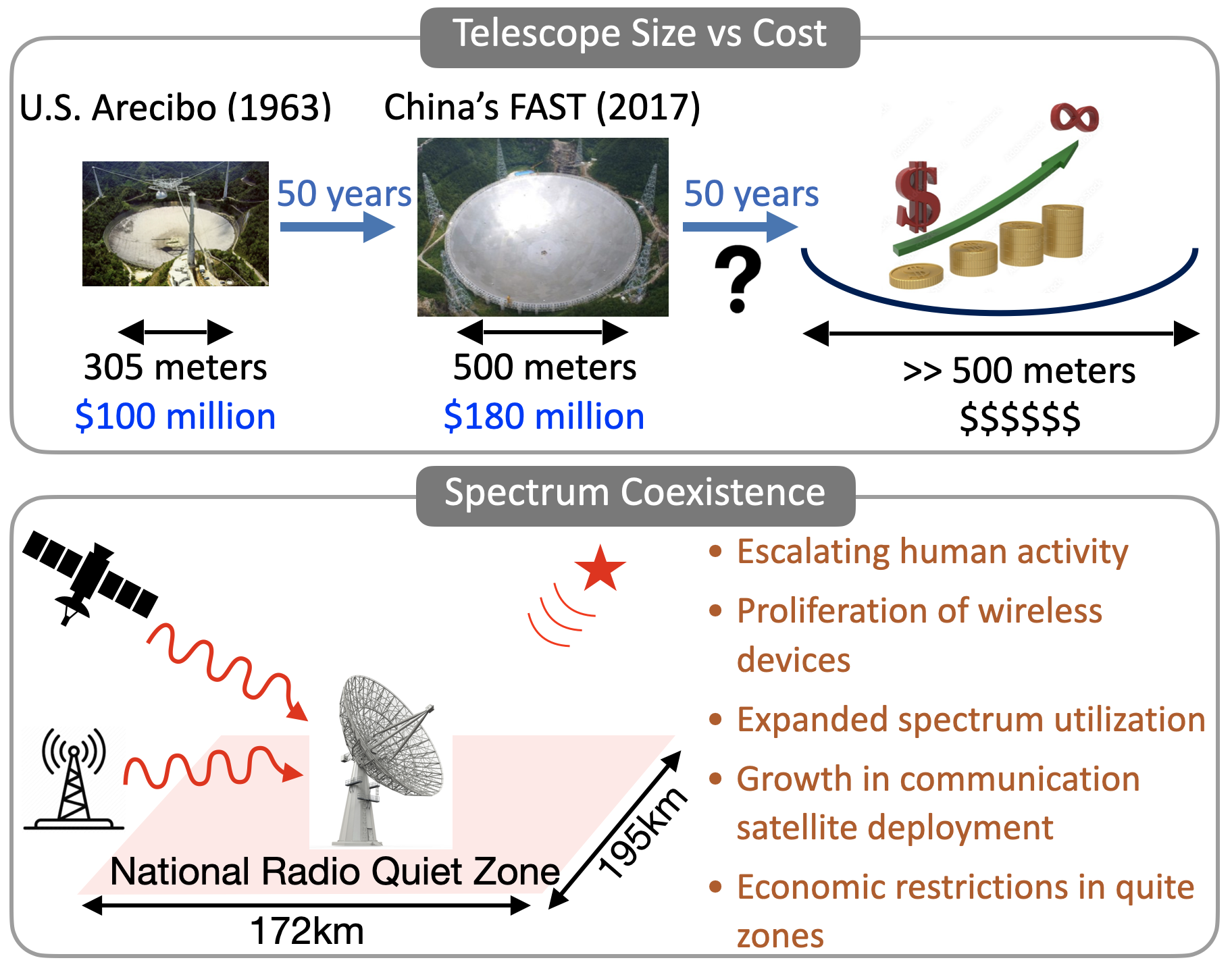}
\caption{Challenges faced by current radio telescopes.}
\label{fig:current_telescope_issue}
\end{figure}

\textit{Challenges for Current Radio Telescopes:} Maintaining a high signal-to-noise ratio (SNR) amidst faint cosmic radio signals, often ranging from $-60$ dB to $-30$ dB, necessitates large reflector antennas and highly sensitive detection systems. As shown in Fig.~\ref{fig:current_telescope_issue}, this requirement has led to the construction of major observatories, such as the 305-meter Arecibo radio telescope, built in 1963 at an inflation-adjusted cost of approximately \$100 million~\cite{Campbell2013AreciboFifty}, and China's 500-meter Aperture Spherical Telescope (FAST), completed in 2017 for around \$180 million~\cite{Cyranoski2018Nature}. These projects reflect a trend toward larger astronomical instruments; however, costs increase exponentially with the size of the reflector antenna~\cite{hansen1981fundamental, davis2011fundamental, geyi2003physical}.

Furthermore, modern wireless communication networks, which utilize an increasing number of spectrum bands, pose significant challenges to the operations of passive radio telescopes. To protect facilities like the Green Bank Observatory, the United States established the National Radio Quiet Zone (NRQZ)~\cite{kellermann2020nrao, abidin2021radio, kellermann2020largest}, encompassing 13,000 square miles. However, the proliferation of wireless devices and smart city technologies near NRQZ boundaries in Fig.~\ref{fig:current_telescope_issue} could lead to increased interference~\cite{beaudet2013radio, abidin2021radio, kellermann2020nrao, sizemore1991national}. Additionally, economic development near radio quiet zones is restricted to preserve astronomical observations, limiting local economic opportunities.

\begin{figure*}[htp]
\centering
\includegraphics[width=0.85\textwidth]{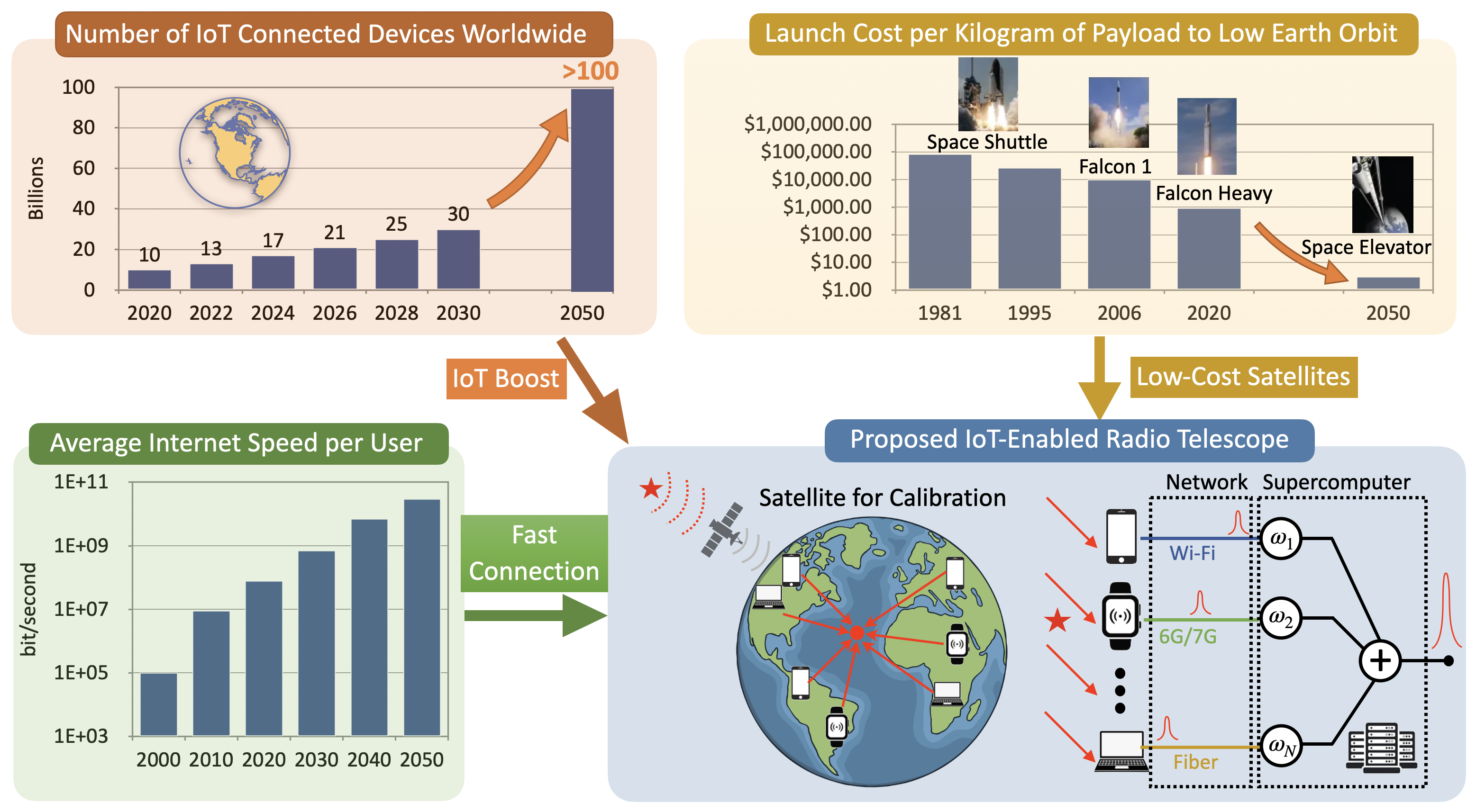}\vspace{-3mm}
\caption{The proposed IoT radio telescope concept.}
\label{fig:motivation}
\end{figure*}

\textit{Proposed Solution:} This paper presents a novel concept that harnesses the global deployment of Internet of Things (IoT) devices to construct a revolutionary radio telescope, with satellites providing calibration. By integrating seamlessly with current IoT infrastructure and communication networks—with only minimal modifications—the system operates using the same frequency, time, and location as existing communication signals, thereby eliminating the need for additional hardware or bandwidth. The idea is inspired by significant technological and economic trends that are set to transform both communication and astronomical observation, as illustrated in Fig.~\ref{fig:motivation}. (1) \textit{IoT Expansion:} A rapidly growing number of IoT devices underpins this proposal; forecasts predict that the global tally of connected IoT devices could surpass 100 billion by 2050~\cite{karunarathne2018wireless}, offering a vast, distributed sensor network that can collectively function as a large-scale radio telescope. (2) \textit{Advancements in Connectivity:} Driven by breakthroughs in future networks and advanced fiber optics, the average global internet speed per user is expected to exceed 30 Gbps by 2050~\cite{FutureTimeline2050, ITU_M2160}, facilitating the swift transfer of extensive observational data to centralized data centers and supercomputing facilities. (3) \textit{Reduced Satellite Costs:} In addition, the significant reduction in satellite launch costs into low Earth orbit, from the era of the Space Shuttle to emerging technologies such as space elevators by 2050, has made space-based platforms increasingly accessible. This advancement enables the deployment of specialized calibration satellites that not only enhance system calibration but also considerably expand the field of view of the IoT radio telescope. Furthermore, for a low cost consideration, the commercial low-orbit communication satellites may also be utilized as calibration sources.

\begin{figure*}[htp]
\centering
\includegraphics[width=0.95\textwidth]{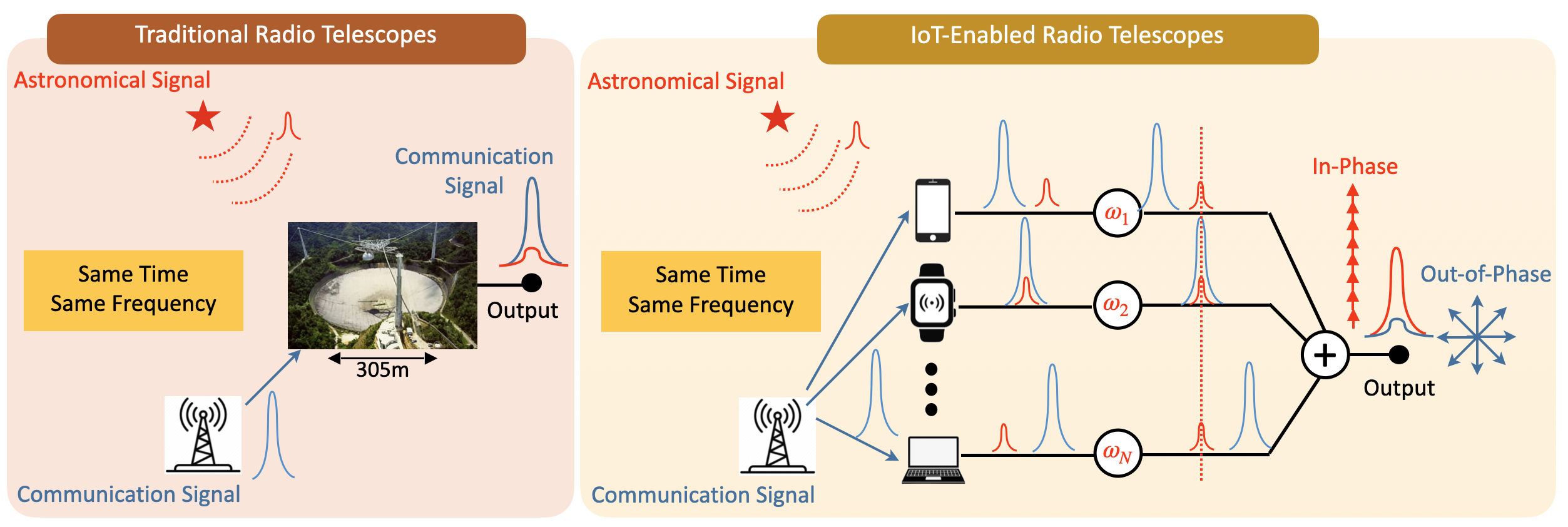}
\caption{Enabled spectrum sharing: (left) traditional radio telescope limitations; (right) proposed IoT-enabled telescope facilitating spectrum sharing.}
\label{fig:spectrum_sharing}
\end{figure*}

\textit{Spectrum Sharing:} Traditional radio telescopes encounter substantial challenges in spectrum sharing primarily due to their structural design and operational environments~\cite{jaroenjittichai2017radio, baan2004radio, bolli2013mobile, briggs2000removing, kesteven2007radio}. As shown in Fig.~\ref{fig:spectrum_sharing} (left), these instruments simultaneously capture both astronomical and wireless communication signals at the same frequencies. Due to the physical limitations of using a solid, large single reflector antenna system, they lack the ability to process or differentiate these signals independently~\cite{leshem2000multichannel, briggs2005overview, fridman2001rfi, bhat2005radio, an2017radio, joardar2007design}. Consequently, communication signals often disrupt astronomical observations, introducing noise that can obscure or distort valuable scientific data~\cite{lanzerotti2001space, penzias1973millimeter, leshem2000radio, national2010spectrum, bean2022casa, emrich1993suggestions, dai2019spectrum}.

\textcolor{black}{The proposed radio telescope design depicted in Fig.~\ref{fig:spectrum_sharing} (right) overcomes these challenges by leveraging a network of widely dispersed IoT devices. Each device is configured to simultaneously capture astronomical and communication signals at the same frequency, time, and location. Distributed digital beamforming technique~\cite{mghabghab2020open, nanzer2021distributed, peng2015radio} is then applied to phase-align the astronomical signals, thereby maximizing the system's output SNR. Moreover, as astronomical signals typically lie below the noise floor, mitigating strong communication interference renders the residual received signal approximately noise-limited. This processing strategy can treat communications signals as background noise and suppress them without corrupting astronomical signals.}

\textcolor{black}{This means the proposed IoT-telescope can simultaneously support communication services (e.g., Wi-Fi, cellular) and contribute to astronomical observations, even when operating within the same frequency bands. }

\begin{figure}[htp]
\centering
\includegraphics[width=0.48\textwidth]{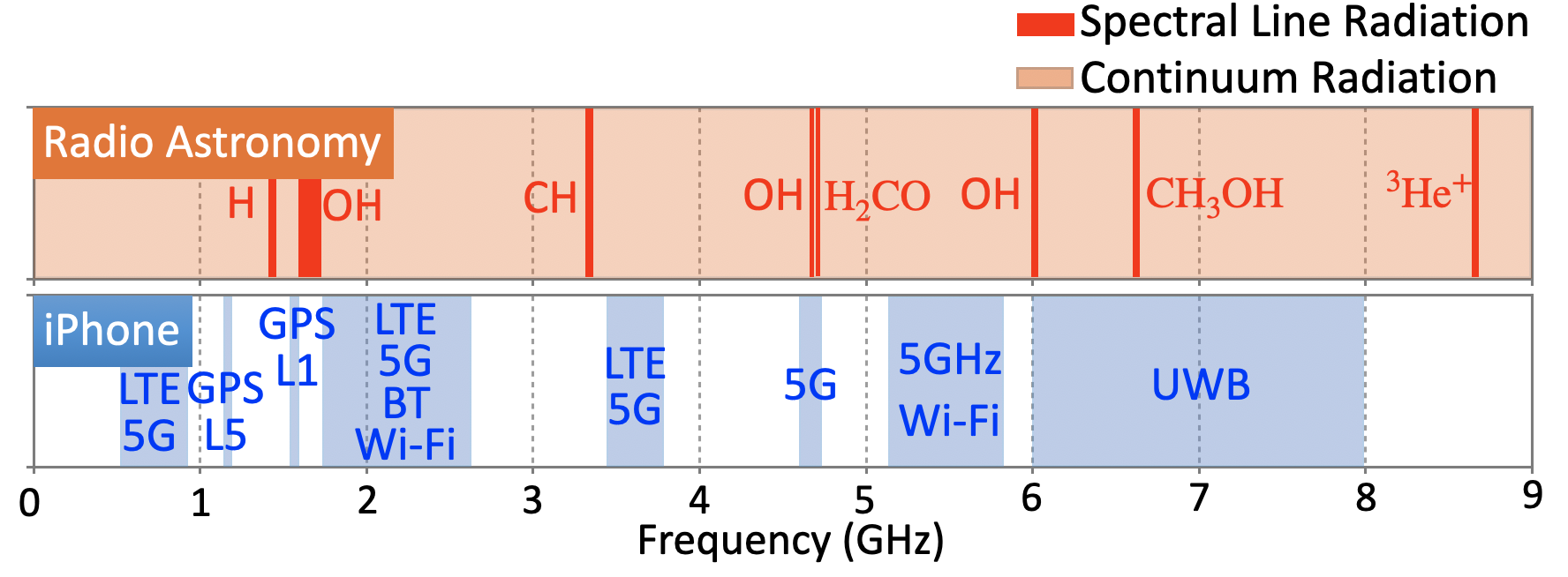}
\caption{Frequency bands of interest for radio astronomy and commercial cell phones below 9 GHz.}
\label{fig:spectrum_phone_Astronomy}
\end{figure}

\textcolor{black}{\textit{Dual-Use IoT Hardware:}} \textcolor{black}{Fig.~\ref{fig:spectrum_phone_Astronomy} shows the frequency bands below 9~GHz that are relevant to both radio astronomy and commercial cellular systems. While radio astronomy seeks to detect both wideband continuum emissions and narrowband spectral lines from atomic and molecular transitions~\cite{thompson2017interferometry}, many modern IoT devices, such as the iPhone 16~\cite{appleStore}, already operate across wide frequency ranges and support multiple bands that overlap with those used in radio astronomical observations. These wideband RF chains, along with high-speed ADCs and onboard processors for IoT devices, suggest a promising opportunity for dual-use spectrum and hardware. In addition, to increase compatibility with key astronomical bands such as the protected 1.4~GHz hydrogen line, future standards could require manufacturers to include additional sensing support for radio astronomy. The cost of these enhancements could be offset by auctioning shared access to protected spectrum bands at 1.4 GHz for radio astronomy, or by reallocating funding from traditional telescope construction, since the IoT telescope relies on widely available distributed and low-cost infrastructure.}

\textit{Significance:} The capability for efficient spectrum sharing between astronomical and communication signals unlocks new avenues for radio astronomy. It enables radio telescopes to operate in densely populated or electronically active regions—areas that were previously unsuitable due to the need for remote, radio-quiet sites—while leveraging the ubiquitous IoT infrastructure to reduce reliance on costly, dedicated facilities. Furthermore, by allowing participants to select their preferred level of involvement, this approach broadens the citizen science community. Overall, the incentive scheme not only improves network scalability and encourages public collaboration in scientific discovery, but it also has the potential to boost public interest in science and technology, create valuable educational opportunities, and promote Science, Technology, Engineering, and Mathematics (STEM) engagement across diverse populations.

\textit{High Performance Prediction:} In comparison to the FAST, the IoT-enabled telescope achieves a two orders-of-magnitude increase in antenna gain and a seven orders-of-magnitude improvement in survey speed. This dramatic enhancement results from the vast number of nodes and the full spherical field of view (FoV). These outcomes highlight the practicality and substantial benefits of the IoT-enabled design, paving the way for cost-effective, rapid, and widely accessible astronomical observations.

The remainder of this paper is organized as follows. Section II describes the system architecture and operational modes of the proposed IoT radio telescope. Section III reviews the principal challenges and their corresponding solutions. Section IV details the calibration methods utilizing both single and multiple satellites. In Section V, we analyze the calibration approaches and assess system performance under various timing accuracy and interference conditions. Section VI provides a comparative evaluation with the FAST. Finally, Section VII concludes the paper.

\section{System Overview}

In this section, we present an overview of the proposed radio telescope system that leverages the global network of IoT devices. The system is designed for seamless integration with existing IoT infrastructure and communication networks, requiring only minimal modifications.

\subsection{IoT System Architecture}
\begin{figure*}[htp]
\centering
\includegraphics[width=0.9\textwidth]{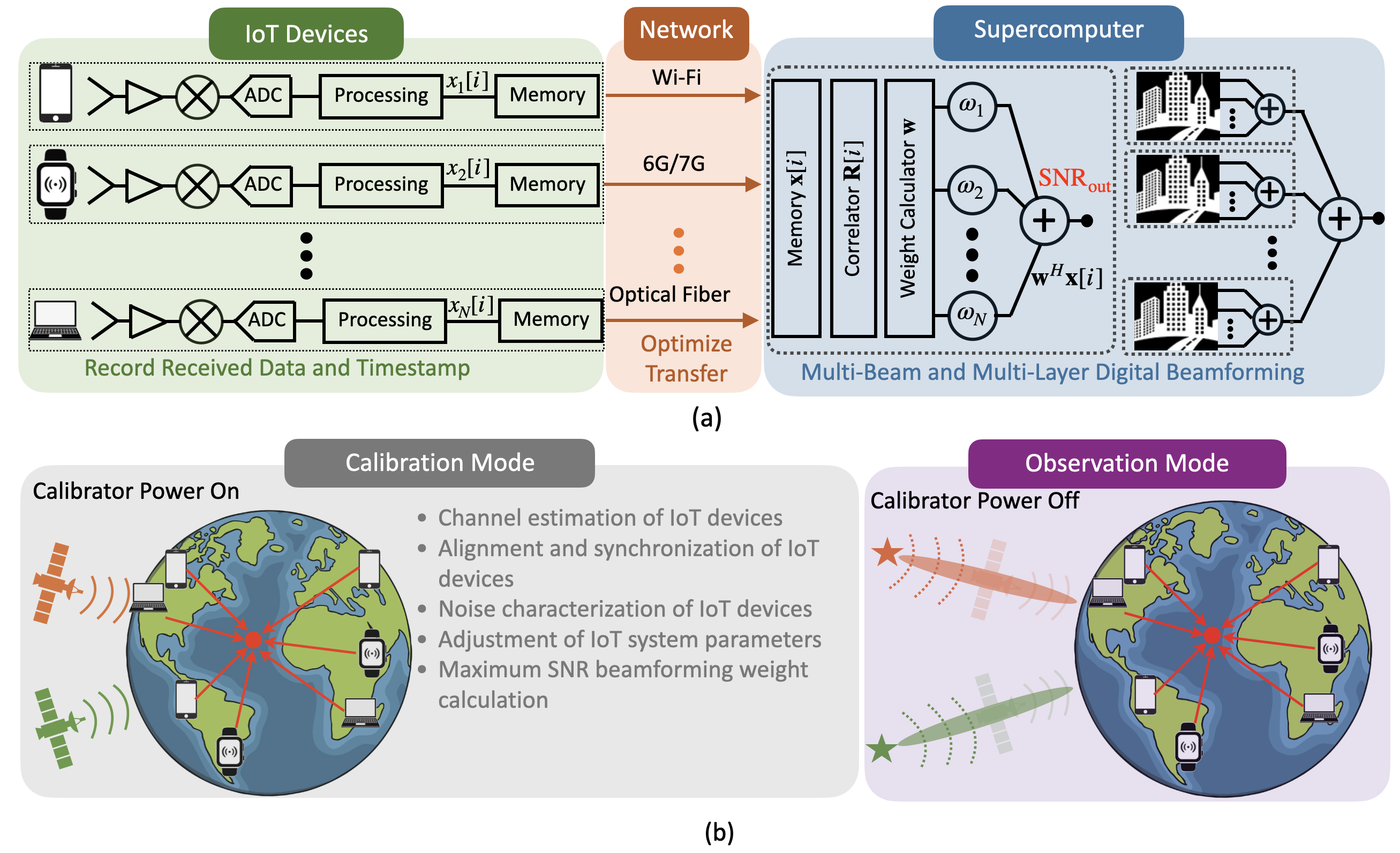}
\caption{Overview of the proposed system: (a) system architecture; (b) calibration and observation modes.}
\label{fig:system_overview}
\end{figure*}

\textcolor{black}{Fig.~\ref{fig:system_overview}(a) illustrates the system architecture, where typical IoT devices are equipped with an antenna, low noise amplifier (LNA), mixer, ADC, processing unit, and memory. Many commercial devices, such as smartphones and smart gateways, already possess these components. For IoT devices that lack full support, minimal updates can be made, such as enabling raw I/Q sample access via firmware, adding a low-cost broadband antenna to extend frequency coverage, and integrating a low-power GPS-disciplined oscillator (GPSDO) for timing coherence. A lightweight software module handles calibration scheduling, local data processing, and spectral data forwarding, all while maintaining the device’s primary IoT functionality.}

\subsection{Data Acquisition and Transmission}

At each time sample $i$, the data vector $\mathbf{x}[i]$ collected by the IoT devices is represented as~\cite{jeffs2008signal}:
\begin{align}
\mathbf{x}[i] = \mathbf{a}\, s[i] + \sum_{q = 1}^{Q} \mathbf{v}_{q}[i]\, d_{q}[i] + \mathbf{n}[i],
\end{align}
where $s[i]$ is the astronomical signal of interest, $d_{q}[i]$ represents one of $Q$ interfering communication signals, $\mathbf{n}[i]$ denotes the noise vector, and $\mathbf{a}$ and $\mathbf{v}_{q}[i]$ are normalized array responses to unit amplitude point sources in the far field corresponding to $s[i]$ and $d_{q}[i]$, respectively.

The data $\mathbf{x}[i]$ can be efficiently transmitted over the communication network to a centralized data center and supercomputing facility. For non-real-time observations, transmission times can be strategically scheduled during periods when devices are typically idle or charging, such as night-time hours.

\subsection{Signal Processing and Beamforming}

The supercomputing facility computes the time-average correlation matrix $\mathbf{R}$ using $N$ samples:
\begin{align}
\mathbf{R} = \frac{1}{N} \sum_{i = 0}^{N-1} \mathbf{x}[i]\, \mathbf{x}[i]^{H},
\end{align}
where $^{H}$ denotes the Hermitian transpose. Using this correlation matrix, the beamforming weight vector $\mathbf{w}$ is calculated in the direction of the astronomical signal $s[i]$. The combined output signal is then obtained by:
\begin{align}
y[i] = \mathbf{w}^{H} \mathbf{x}[i].
\end{align}

\subsection{Calibration and Observation Modes}
\label{section:Calibration_and_Observation}
As depicted in Fig.~\ref{fig:system_overview}(b), the proposed system operates in two primary modes: \textit{calibration} and \textit{observation}.

\subsubsection{Calibration Mode}

In the calibration mode, specialized satellites act as calibrators by transmitting known calibration signals. These signals are received by the globally distributed IoT devices, enabling precise estimation of channel properties such as time delays and frequency shifts caused by the varied locations and operational environments of the devices. This process ensures synchronization and coherence of the data collected from different devices.

The known calibration signals also aid in characterizing and quantifying noise levels and interference patterns in the received data, enhancing the clarity and accuracy of the subsequent astronomical observations. Calibration facilitates the fine-tuning of each IoT device's parameters, such as gain settings and filter thresholds, thereby maximizing the sensitivity of the receiver system.

The calibration data and timestamps are used to compute the maximum SNR beamforming weight vector $\mathbf{w}_{\text{max,SNR}}$ by solving the generalized eigenvalue problem:
\begin{align}
\mathbf{R}_{\text{sat}}\, \mathbf{w} = \lambda_{\text{max}}\, \mathbf{R}_{n}\, \mathbf{w},
\end{align}
where $\mathbf{R}_{\text{sat}}$ is the correlation matrix of the received calibration signal from the satellite, $\mathbf{R}_{n}$ is the noise correlation matrix (equal to $\mathbf{R}_{\text{off}}$ when the calibration source is off), and $\lambda_{\text{max}}$ is the maximum eigenvalue.

\begin{figure}[htp]
\centering
\includegraphics[width=0.48\textwidth]{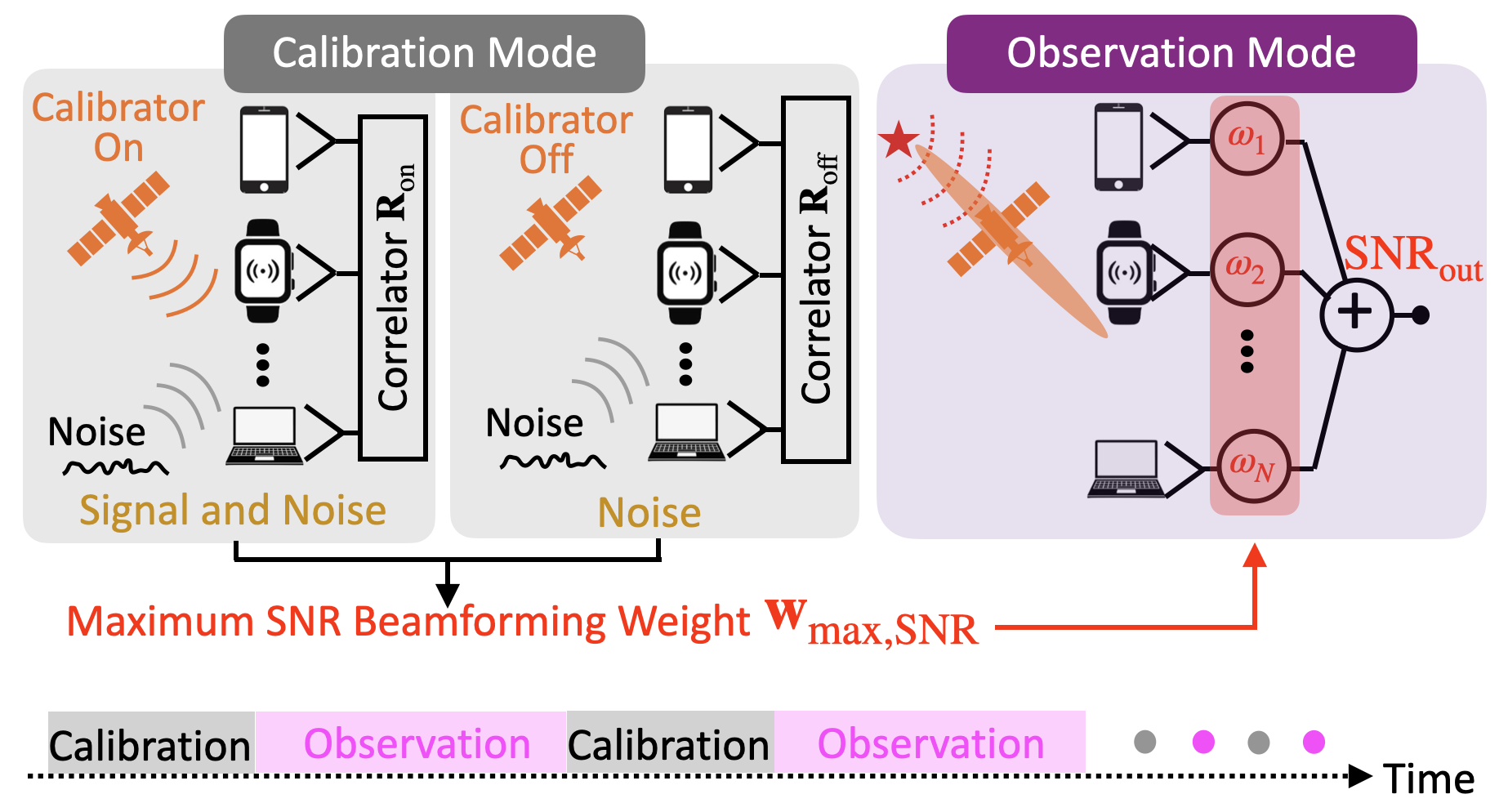}
\caption{Signal processing in calibration and observation modes.}
\label{fig:calibration_description}
\end{figure}

As shown in Fig.~\ref{fig:calibration_description}, $\mathbf{R}_{n}$ is obtained during periods when the calibration source is powered off, and $\mathbf{R}_{\text{sat}}$ is calculated by subtracting $\mathbf{R}_{n}$ from the total received correlation matrix when the calibration source is on:
\begin{align}
\mathbf{R}_{\text{sat}} = \mathbf{R}_{\text{on}} - \mathbf{R}_{\text{off}}.
\end{align}

\subsubsection{Observation Mode}

Once the optimal beamforming weights $\mathbf{w}_{\text{max,SNR}}$ are determined, the system transitions to observation mode to detect astronomical signals. The output SNR in observation mode is expressed as:
\begin{align}
\text{SNR}_{\text{out}} = \frac{\mathbf{w}_{\text{max,SNR}}^{H} \mathbf{R}_{\text{sig}}\, \mathbf{w}_{\text{max,SNR}}}{\mathbf{w}_{\text{max,SNR}}^{H} \mathbf{R}_{n}\, \mathbf{w}_{\text{max,SNR}}},
\end{align}
where $\mathbf{R}_{\text{sig}}$ is the correlation matrix of the astronomical signal.

Due to variations in signal and noise caused by environmental factors, the system periodically alternates between calibration and observation modes to maintain accuracy and coherence in data collection.

\section{Overview of Challenges and Solutions}
Despite the promise of leveraging ubiquitous IoT devices as a distributed radio telescope, several technical challenges must be overcome to achieve coherent, high-fidelity operation at scale. Table.~\ref{fig:challenges_solutions} provides an overview of these key challenges and their corresponding solutions. In this section, each challenge is examined in detail along with the engineering solutions employed to address them.

\begin{table*}[h]
\caption{Summary of technical challenges and solutions.}
\centering
\includegraphics[width=0.85\textwidth]{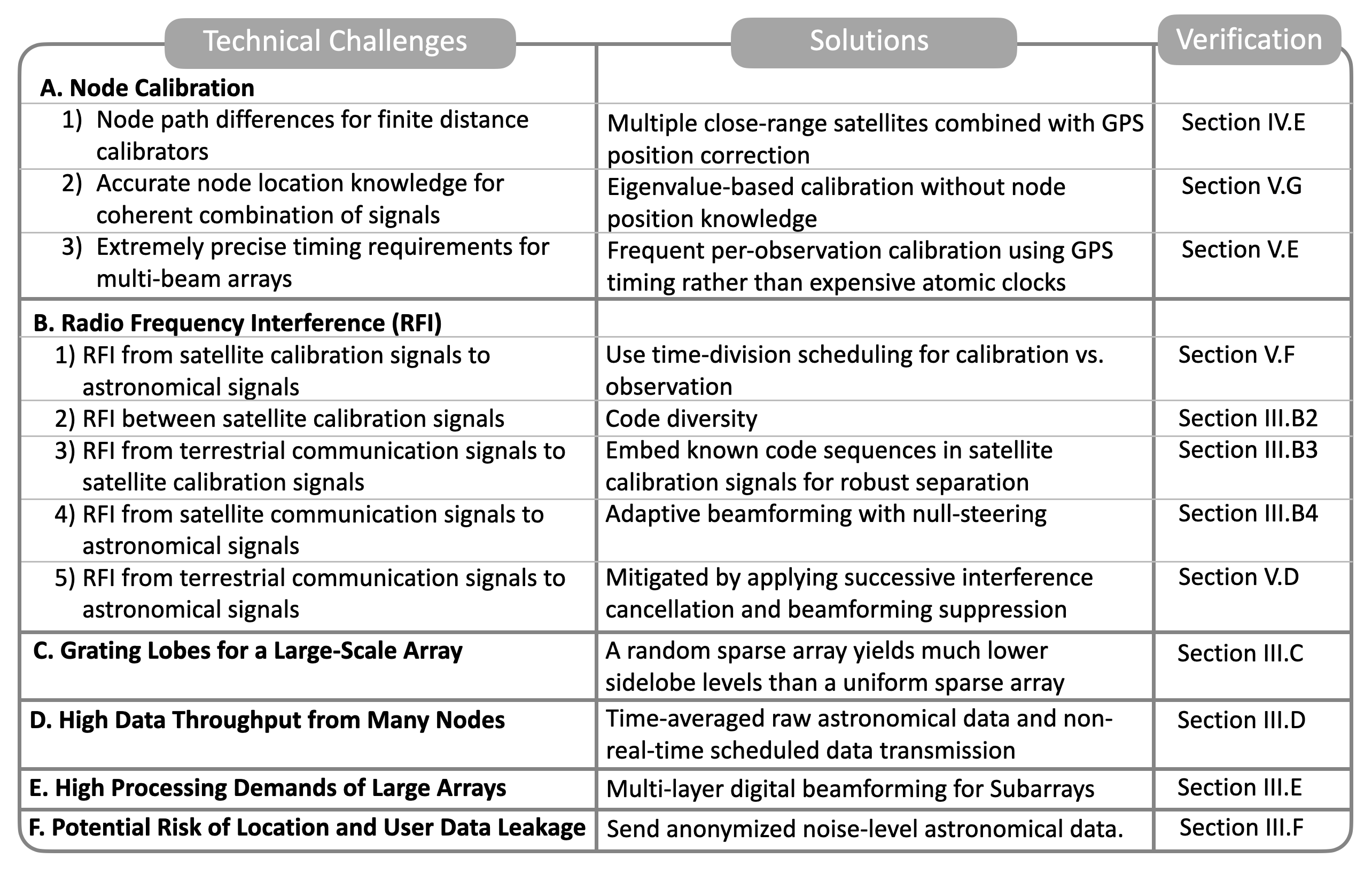}
\label{fig:challenges_solutions}
\end{table*}

\subsection{Node Calibration}
\subsubsection{Node Path Differences for Finite Distance Calibrators}
\begin{figure}[htp]
\centering
\includegraphics[width=0.45\textwidth]{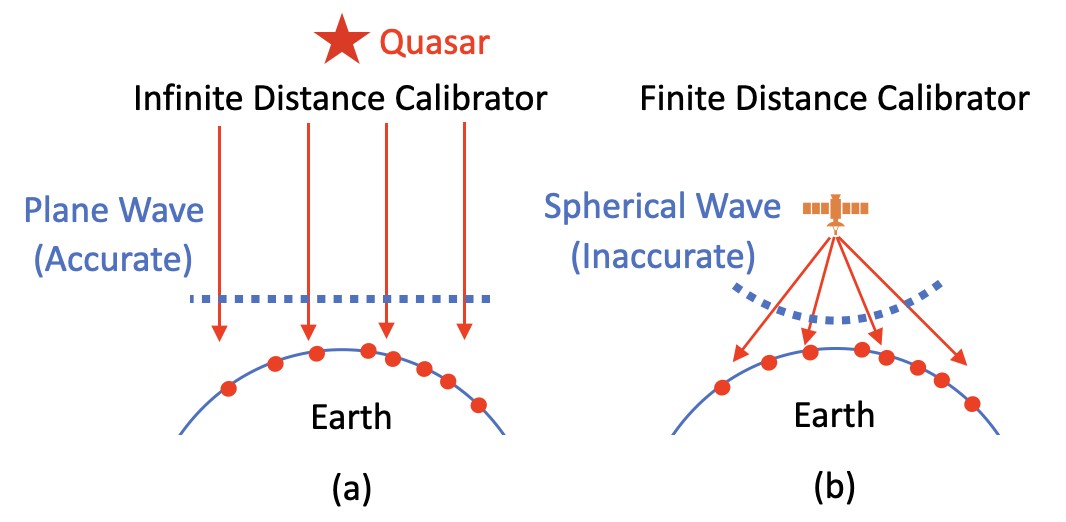}
\caption{Infinite distance calibrator vs. finite distance calibrator.}
\label{fig:distance_calibrator}
\end{figure}

Traditional interferometer radio telescopes are calibrated using well‐characterized astronomical sources, such as quasars, which are effectively at infinite distance. As described in Fig.~\ref{fig:distance_calibrator}(a), the calibrator’s wavefront is nearly a perfect plane wave across the telescope’s aperture. However, when a satellite is used as a calibration source in Fig.~\ref{fig:distance_calibrator}(b), its finite distance results in a curved wavefront, which introduces phase discrepancies among the nodes relative to the ideal plane-wave condition.

\textcolor{black}{To mitigate this issue, we adopt a multi-satellite, GPS-aided calibration strategy. Each calibration satellite is assigned to a geographically bounded subarray so that, within that subregion, the spherical wavefront is approximated by a plane wave. In addition, GPS/GNSS-based localization of node position enables geometric pre-correction that removes residual curvature-induced phase terms \emph{prior} to calibration. With this pre-correction, the calibration under its plane-wave assumption; without it, the number of satellites required to maintain local planarity increases sharply (see Fig.~\ref{fig:satB_diff_GPSError}). The details of this method and the calculation results are described in Section~\ref{section:multiple_sat_calibration}.}

\subsubsection{Accurate Node Location Knowledge}
For a traditional interferometry array, the position of each array node must be known to within a fraction of the wavelength to ensure coherent signal combination. Even minor positional errors can cause significant phase misalignments, resulting in signal cancellation and reduced sensitivity. In contrast, the IoT-enabled telescope utilizes an eigenvalue-based calibration method that generates an effective beam in the calibrator's direction without requiring precise knowledge of each node's position. The details of the method and the calculation results are described in Section~\ref{section:system_signal_model}.

\subsubsection{Extremely Precise Timing Requirements for Multi-Beam Arrays}
Traditional interferometry arrays employ hydrogen maser atomic clocks at each station to achieve synchronization at the femtosecond level~\cite{bandi2024comprehensive}. In contrast, IoT nodes rely on GPS timing, which can provides \textless1 nanosecond accuracy using GPSDO~\cite{ettusUB210}. To compensate for the lower timing precision, IoT nodes perform frequent per-observation frequency calibration. As described in Section~\ref{section:system_signal_model}, provided that if the calibration is performed frequently enough, the frequency drift of the node clocks can be effectively mitigated before it significantly degrades coherence of the IoT radio telescope.

\subsection{Radio Frequency Interference (RFI)}
\subsubsection{RFI from Satellite Calibration Signals to Astronomical Signals}
The calibration signals transmitted by satellites could inadvertently overwhelm the weak cosmic signals that the array nodes are designed to detect. As illustrated in Fig.~\ref{fig:calibration_description}, a time-division scheduling scheme isolates this interference. The simulation model described in Section~\ref{section:system_signal_model} implements this approach by having satellites transmit known calibration signals during brief calibration intervals. During these intervals, nodes measure and update their phase offsets while refraining from collecting astronomical data. In the subsequent observation period, the satellites cease transmission, allowing nodes to record only natural cosmic signals with the calibration already applied. By rapidly toggling between calibration and observation, the system ensures that the strong satellite calibration signals do not drown out the astronomical signals.

\textcolor{black}{\subsubsection{RFI Between Satellite Calibration Signals}
As illustrated in Fig.~\ref{fig:schematic_satB}, multiple distributed calibration satellites are employed to address the curvature of the wavefront introduced by the finite distance of each calibrator. This multi-satellite setup not only compensates for near-field effects but also enables full-sphere field-of-view coverage through multi-beam digital beamforming. To prevent interference among satellites transmitting on the same observation frequency band, we implement code diversity by assigning each satellite a unique orthogonal spreading code, such as Gold or Zadoff-Chu sequences. This approach allows simultaneous transmission across the constellation without cross-interference. In regions where coverage zones overlap, IoT receivers correlate incoming signals with all known codes, allowing them to distinguish and coherently combine data from multiple calibration sources.}

\subsubsection{RFI from Terrestrial Communication Signals to Satellite Calibration Signals}
Satellite calibration signals are encoded with distinct coded sequences that serve as identifiable signatures. IoT receivers leverage these known codes to accurately detect and isolate the calibration signals from any overlapping terrestrial communication signals.

\subsubsection{RFI from Satellite Communication Signals to Astronomical Signals}
For predictable external RFI stemming from ground-to-satellite communications, dynamic beamforming with null-steering is employed~\cite{jeffs2005auxiliary, sardarabadi2015spatial}. The array’s beamforming algorithms adaptively place nulls in the direction of known interfering satellites, effectively suppressing their signals.

\subsubsection{RFI from Terrestrial Communication Signals to Astronomical Signals}\label{sec:SIC_Inter_Mitigation}
The proposed system employs successive interference cancellation (SIC) to mitigate strong terrestrial communication signals prior to distributed beamforming and correlation. In this approach, dominant interference components are first identified and estimated using their known or partially known signal structures, and are subsequently subtracted from the received waveform. Since astronomical signals typically lie well below the thermal noise floor, they are not directly affected by this cancellation process. The primary objective of SIC is to suppress strong interference and, more importantly, to reduce its spatial coherence across the distributed IoT nodes, ensuring that any residual interference combines incoherently in the beamformer.

By mitigating interference through SIC, the system preserves the statistical properties of the noise-dominated astronomical signal while maintaining linear operation of the receiver front-end. This enables the telescope array to operate concurrently with terrestrial communication systems within shared frequency bands. As a result, IoT devices can continue performing their primary communication functions (e.g., Wi-Fi, cellular) while simultaneously contributing data for astronomical observations. The implementation details and numerical evaluation of the proposed SIC-based approach are presented in Section~V.

\subsection{Grating Lobes for a Large-Scale Array}
Grating lobes typically appear in uniform arrays when the element spacing exceeds half the wavelength. However, for a large-scale random array, the sidelobe region of the radiation pattern consists of a sum of random phasors due to the irregular distribution of array elements. This randomness leads to a significant reduction in the peak sidelobe level compared to that of a uniform array. As noted in classical work by Y. T. Lo~\cite{lo1964mathematical}, consider a square array operating at 300 GHz with an aperture equivalent to the Earth's diameter and a desired peak sidelobe level (PSLL) of -30 dB. For a random array, approximately $5\times10^{5}$ elements are sufficient, whereas a uniform array with half-wavelength spacing would require about $2\times10^{20}$ elements. Moreover, according to~\cite{diao2017sidelobe}, the PSLL can be approximately calculated by
\begin{equation}
    \text{PSLL (dB)} = -10 \log_{10} \left( D_{\text{ele}} \frac{N_{\text{ele}}}{L/\lambda} \right),
\end{equation}
where $D_{\text{ele}}$ represents the element directivity in the mainlobe direction, $N_{\text{ele}}$ is the number of array elements, and $L$ is the array length. Assuming $D_{\text{ele}}=1$, $N_{\text{ele}}=5\times10^{10}$, $L=1\times10^{7}$ meters, and an operating frequency of 1.42 GHz, the PSLL is approximately -30 dB. Therefore, this rough estimation indicates that grating lobes are unlikely to occur, and the sidelobe levels in the large-scale IoT-enabled telescope can be maintained at very low levels due to the random distribution of nodes.

\subsection{High Data Throughput from Many Nodes}

\textcolor{black}{Aggregating signals from a very large number of IoT nodes inevitably produces substantial data volumes. As a representative example, consider a single node operating over an observation bandwidth of $B = 1$\,GHz with 8\,-bit complex sampling (in-phase and quadrature). Using a Nyquist sampling rate of $f_s = 2B = 2$\,Gsps, the corresponding raw data rate per node is
\begin{equation}
R_{\mathrm{raw}} = f_s \times (\text{bits per sample}) \times 2 \ (\text{I/Q}),
\end{equation}
which for $f_s = 2$\,Gsps and 8-bit I/Q sampling yields
\begin{equation}
R_{\mathrm{raw}} = 2~\mathrm{Gsps} \times 8~\mathrm{bits} \times 2 = 32~\mathrm{Gbps~per~node}.
\end{equation}
As shown in Fig.~\ref{fig:motivation}, average per-user internet speeds are projected to exceed $30$\,Gbps by 2050, making such per-node throughputs technically feasible. However, scaling to hundreds of millions or billions of nodes would push aggregate throughput far beyond the capacity of centralized systems, necessitating strategies that substantially reduce the transmitted data volume.}

\subsubsection{\textcolor{black}{Local Preprocessing to Reduce Data Volume}}
{\textcolor{black}{Transmitting raw streams is neither necessary nor efficient. Each IoT node can perform local preprocessing prior to uplink, including time averaging and data compression. In practice, these steps can reduce data rates by orders of magnitude while preserving the astronomical information of interest~\cite{wijnholds2018baseline,de2025unlocking}. }

\subsubsection{\textcolor{black}{Network Scheduling and Heterogeneous Paths}}
\textcolor{black}{With consumer access speeds expected to surpass $30$\,Gbps per user by 2050 (Fig.~\ref{fig:motivation}), partially processed high-bandwidth datasets can be delivered in near real time. For non-real-time observing modes, transfers can be deferred to off-peak periods to reduce congestion and operational cost. The inherently distributed nature of IoT connectivity further allows load balancing across heterogeneous infrastructures (fiber, 5G/6G, and satellite links).}

\subsubsection{\textcolor{black}{On-Device Storage for Buffered Transfer}}
\textcolor{black}{Many radio-astronomical observations do not require real-time data delivery; consequently, on-device storage can serve as a buffer that decouples acquisition from transmission. Storage technology has advanced rapidly: for example, the first-generation iPhone (2007) had $8$\,GB of storage, whereas the iPhone~17 (2025) offers up to $2$\,TB. Following this trajectory, compact and affordable multi-terabyte or even petabyte-scale storage is likely within the coming decades. This trend would make long-term buffering of large astronomical datasets feasible, supporting asynchronous upload strategies. In practice, nodes record calibration and observation data to on-device storage and transfer it opportunistically (e.g., when idle, on Wi-Fi, or while charging), thereby decoupling acquisition from transmission and minimizing power impact for the low-power IoT devices.}

\subsection{High Computational Processing Requirements}
To enhance the efficiency of correlation and beamforming calculations in a large-scale array, a multi-layer digital beamforming strategy is employed, as shown in Fig.~\ref{fig:system_overview} (a) right. In this scheme, the global IoT network is divided into several smaller regions, or subarrays, where initial beamforming weights are computed over the half spherical scanning range for each local subarray. The locally beamformed signals are subsequently combined and processed through an additional layer of digital beamforming. This hierarchical approach optimizes the distribution of computations, reduces the processing load on individual nodes, and minimizes the required data transmission bandwidth, thereby enhancing overall processing efficiency and scalability.

\subsection{Potential Risk of Location and User Data Leakage}

Integrating personal IoT devices into a large-scale scientific sensing network naturally raises concerns regarding data privacy and security, particularly when devices may be associated with user location or other sensitive information. To mitigate these risks, the proposed framework is designed such that raw communication data or user-specific information is neither required nor transmitted to the central processor.

As discussed earlier, most astronomical signals of interest lie well below the thermal noise floor of the IoT receiver. Furthermore, the successive interference cancellation (SIC) approach introduced in Section~\ref{sec:SIC_Inter_Mitigation} suppresses dominant communication signals prior to distributed beamforming, resulting in residual signals that are effectively noise-limited. Consequently, the data forwarded to the central processor consists primarily of anonymized, noise-level samples that contain no recoverable user communication content or identifiable information.

Under these assumptions, the proposed architecture minimizes the risk of location and user data leakage while preserving the statistical information necessary for astronomical correlation and imaging. This separation between user communication functionality and scientific data processing provides an inherent layer of privacy protection, supporting the safe participation of personal IoT devices in distributed radio astronomy observations.

\subsection{Non-Technical Consideration: Public Engagement and Incentive Plan}

To maximize the operational efficiency of the distributed IoT-based radio telescope network and foster public engagement, an incentive plan should be implemented to encourage individuals to contribute their personal devices—such as smartphones and laptops—as active nodes in the array. This approach transforms everyday technology into valuable resources for astronomical research. Participants would receive compensation through digital credits, service discounts, or monetary payments determined by their contributed processing power and the duration of device participation.

Furthermore, it is advisable to establish a protocol or standard to ensure that future IoT devices are designed to support the frequency and bandwidth requirements essential for radio astronomy, such as the 1.4\,GHz band. The additional costs associated with this enhancement could be offset by auctioning the 1.4\,GHz bandwidth, which is currently reserved for radio astronomy, or by reallocating funds from the construction of new radio telescopes—given that the IoT-enabled telescope leverages nearly free infrastructure.

By allowing participants to determine their level of involvement based on personal preference, this strategy broadens the community of citizen scientists. Overall, the incorporation of such an incentive plan not only enhances network scalability but also fosters public collaboration in scientific discovery. This democratization of astronomical research is expected to boost public interest in science and technology, provide valuable educational opportunities, and promote STEM engagement across diverse populations.

\section{Node Phase Calibration Using Satellites for Simultaneous Multiple Beams}
\label{section:multiple_sat_calibration}
In this section, we introduce a calibration method for large-scale distributed node arrays on Earth's surface utilizing multiple calibration satellites, designated as Satellite B. This technique significantly improves array gain and reduces phase errors compared to methods that rely on a single calibration satellite (Satellite A). By deploying multiple satellites for calibration, the system enables simultaneous beam scanning in various directions, theoretically allowing the generation of an infinite number of beams when integrated with digital beamforming algorithms.

\subsection{System Model and Assumptions}

We consider a large aperture array consisting of $N_{\text{node}}$ nodes randomly distributed along a circular path on Earth's surface. For computational simplicity, we model the array in a two-dimensional plane while accounting for Earth's curvature. The Earth's radius is $R_{\text{earth}} = 6,\!378$\,km.

Each node is assumed to be an isotropic radiator, a common assumption in array antenna theory to simplify analysis \cite{balanis2016antenna}. Due to the large separation distances between nodes, mutual coupling effects are neglected. The array operates at a central frequency of $f = 1.42$\,GHz, corresponding to the resonant frequency of neutral hydrogen, which is significant in radio astronomy. The nodes form a large aperture array with a baseline length $L_{\text{baseline}}$. The steering angle $\theta_{\text{steer}}$ ranges from $-90^\circ$ to $90^\circ$, enabling beam scanning across a broad angular range. 

Two sets of calibration satellites are utilized: Satellite A, consisting of a single calibration satellite located at an altitude of $h_{\text{satA}}$ above Earth's surface, and Satellite B, comprising multiple calibration satellites evenly distributed along an orbital path at an altitude of $h_{\text{satB}}$, with a total number of $N_{\text{satB}}$.

\begin{figure}[htp!]
      \begin{center}
      \includegraphics[scale=0.38]{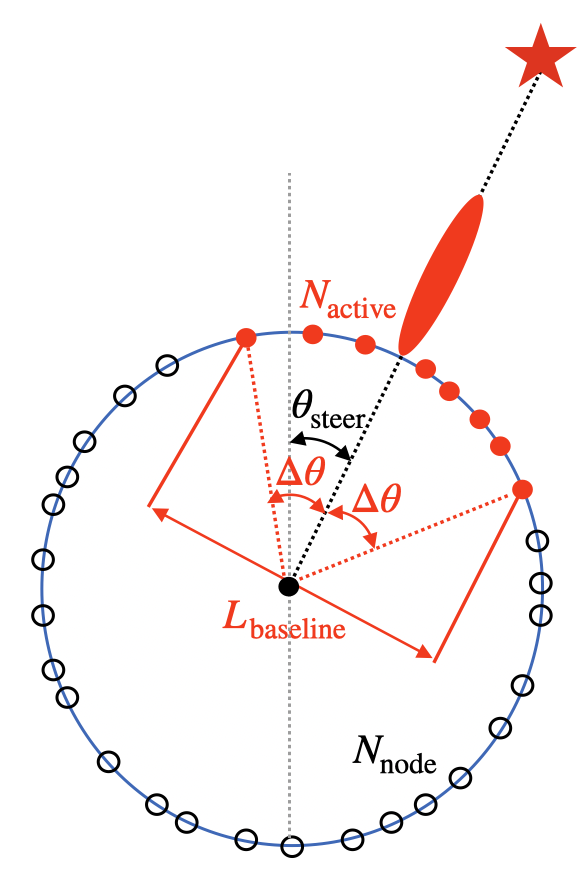}
      \caption{Schematic of active array nodes for beamforming in a 2-D plane.}
      \label{fig:schematic_baseline}
      \end{center}
\end{figure}

\subsection{Ideal Array Gain}

When the beam is steered in the $\theta_{\text{steer}}$ direction, the active nodes for beamforming described in Fig.~\ref{fig:schematic_baseline} are selected within a angle range defined by $\theta_{\text{steer}} \pm \Delta\theta$, where

\begin{equation}
    \Delta\theta = \arcsin\left(\frac{L_{\text{baseline}}}{2 R_{\text{earth}}}\right).
\end{equation}
This ensures that the nodes included in the calibration are within the desired baseline length. The number of active nodes is calculated as

\begin{equation}
    N_{\text{active}} = \left\lfloor \frac{2 \Delta\theta}{2\pi} N_{\text{node}} \right\rfloor.
\end{equation}

Assuming isotropic sources and coherent summation, the ideal array gain is given by

\begin{equation}
    G_{\text{ideal}} = 10 \log_{10} N_{\text{active}}^{2}.
    \label{eq:ideal_gain}
\end{equation}

\subsection{Calibration Using Single Calibration Satellite (Satellite A)}

In this calibration method, as shown in Fig.~\ref{fig:schematic_satA}, a single calibration satellite (Satellite A) is used to send calibration signals to each node. The calibration process involves determining the relative phase delays between nodes based on the received calibration signals.

\begin{figure}[htp!]
      \begin{center}
      \includegraphics[scale=0.36]{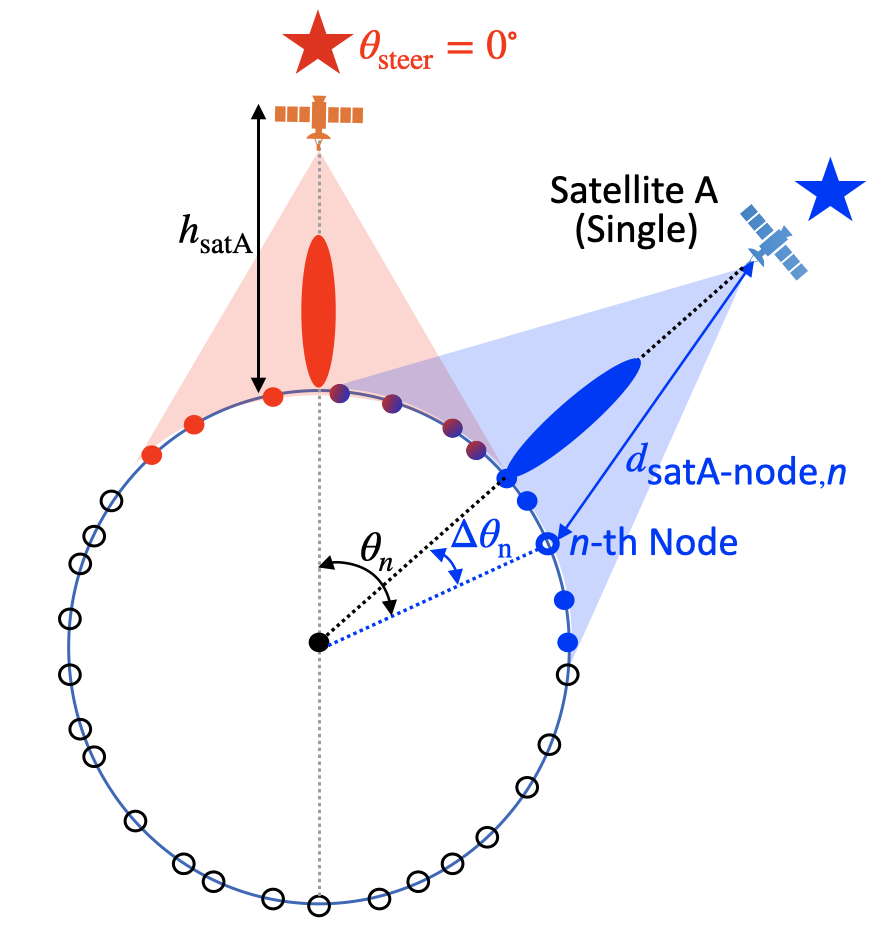}
      \caption{Schematic for node calibration using single calibration satellite (Satellite A).}
      \label{fig:schematic_satA}
      \end{center}
\end{figure}

The distance between Satellite A and the $n$-th node is denoted as $d_{\text{satA-node},n}$, and the corresponding calibration phase shift is calculated by

\begin{equation}
    \phi_{\text{cal,satA},n} = \frac{2\pi}{\lambda} d_{\text{satA-node},n}, \quad n = 1, 2, \dots, N_{\text{active}}.
\end{equation}

We define the angular difference between each node and the steering direction as

\begin{equation}
    \Delta\theta_n = \theta_{\text{steer}} - \theta_n,
\end{equation}
where $\theta_n$ is the angular position of the $n$-th node.

The geometric phase term due to Earth's curvature is given by

\begin{equation}
    \phi_{\text{geom,satA},n} = -\frac{2\pi}{\lambda} R_{\text{earth}} \left(1 - \cos\Delta\theta_n\right).
\end{equation}

Assuming isotropic sources and applying uniform weighting, the array factor after calibration with Satellite A is

\begin{equation}
    AF_{\text{satA}} = \sum_{n=1}^{N_{\text{active}}} \exp\left( j \phi_{\text{geom,satA},n} + j \phi_{\text{cal,satA},n} \right).
    \label{eq:AFsatA}
\end{equation}

The array gain is calculated as the normalized magnitude squared of the array factor:

\begin{equation}
    G_{\text{satA}} = 10 \log_{10}  \left| AF_{\text{satA}} \right|^2 .
\end{equation}

In the far-field region, or when the distance between Satellite A and Earth is large enough, the calibration phase $\phi_{\text{cal,satA},n}$ can be used to well compensate the geometric phase $\phi_{\text{geom,satA},n}$ variation among the nodes. In this case, the array gain $G_{\text{satA}} $ is approximately equal to the ideal array gain $G_{\text{ideal}}$.

This calibration approach relies on a single satellite to receive astronomical signals from a specific direction. To achieve simultaneous multiple scanned beams, multiple satellites like Satellite A would be required. In sub-regions where the coverage of different satellites overlaps, like the overlap of red and blue sub-regions in Fig.~\ref{fig:schematic_satA}, the calibration signals can be separated by employing different frequencies or coding schemes to avoid ambiguity.

\subsection{Calibration Using Multiple Distributed Calibration Satellites (Satellite B)}

To enhance the calibration process, we introduce multiple distributed calibration satellites (Satellite B) in Fig.~\ref{fig:schematic_satB}. Each satellite in Satellite B is responsible for calibrating a specific sub-region of nodes. The direction from the center of each node sub-region to its corresponding satellite is aligned with the beam steering direction.

\begin{figure}[htp!]
      \begin{center}
      \includegraphics[scale=0.36]{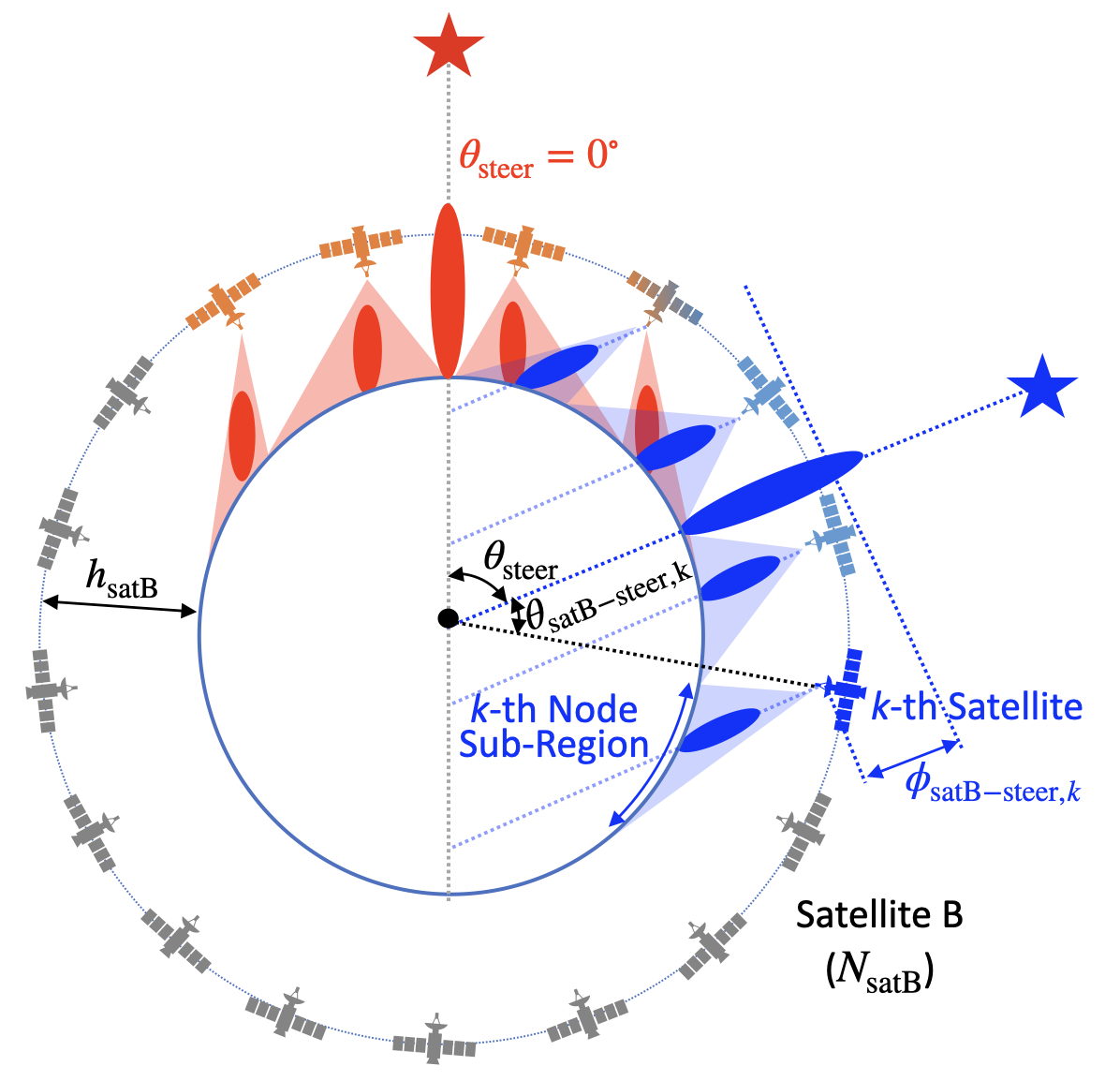}
      \caption{Schematic for node calibration using multiple distributed calibration satellites (Satellite B).}
      \label{fig:schematic_satB}
      \end{center}
\end{figure}

The angular positions of the satellites in Satellite B are defined as
\begin{equation}
    \theta_{\text{satB},k} = -\pi + \left( k - \frac{1}{2} \right) \frac{2\pi}{N_{\text{satB}}}, \quad k = 1, 2, \dots, N_{\text{satB}}.
\end{equation}

For the $n$-th node in the $k$-th sub-region, the calibration phase involves calculating the relative phase variation due different distance $d_{\text{satB-node},n}$ between the node and its corresponding satellite and relative phase $\phi_{\text{satB-steer},k}$ among each satellites. The calibration phase shift is
\begin{equation}
    \phi_{\text{cal,satB},n} = \frac{2\pi}{\lambda} d_{\text{satB-node},n} + \phi_{\text{satB-steer},k}.
\end{equation}
Here, $\phi_{\text{satB-steer},k}$ accounts for the phase difference between the $k$-th satellite and the plane wavefront in the steering direction, calculated by
\begin{equation}
    \phi_{\text{satB-steer},k} = \frac{2\pi}{\lambda} \left( h_{\text{satB}} + R_{\text{earth}} \right) \left( 1 - \cos \theta_{\text{satB-steer},k} \right),
\end{equation}
where $\theta_{\text{satB-steer},k}$ is the angle difference between the $k$-th satellite and the steering direction:
\begin{equation}
    \theta_{\text{satB-steer},k} = \theta_{\text{steer}} - \theta_{\text{satB},k}.
\end{equation}

The geometric phase term for nodes due to Earth's curvature is

\begin{equation}
    \phi_{\text{geom},n} = \frac{2\pi}{\lambda} R_{\text{earth}} \cos \Delta\theta_n.
\end{equation}

Assuming isotropic sources and uniform weighting, the array factor after calibration with Satellite B is

\begin{equation}
    AF_{\text{satB}} = \sum_{k=1}^{N_{\text{satB}}} \sum_{n \in \mathcal{R}_k} \exp\left( j \phi_{\text{geom},n} + j \phi_{\text{cal,satB},n} \right),
    \label{eq:AFsatB}
\end{equation}
where $\mathcal{R}_k$ represents the set of nodes in the $k$-th sub-region associated with the $k$-th satellite.

The array gain is then calculated as

\begin{equation}
    G_{\text{satB}} = 10 \log_{10} \left( \left| AF_{\text{satB}} \right|^2 \right).
\end{equation}

As shown in Fig.~\ref{fig:schematic_satB}, the satellites in Satellite B and the nodes can be multiplexed for different scanning beams. As the beam is steered, the node regions and corresponding satellites can be redefined, allowing for flexible beamforming capabilities.

\subsection{Phase Error Mitigation Using GPS Localization}
Unlike astronomical signals originating from sources at infinite distances, the finite distance of Satellite B introduces phase variations in the wavefront curvature across the array, leading to phase discrepancies among the nodes. To mitigate these phase errors, we employ GPS localization to estimate the positions of both the nodes and the satellites, allowing us to apply appropriate phase corrections during calibration.

Each node and satellite may have position estimation errors due to limitations in GPS accuracy. Let $\Delta_{\text{node,GPS}}$ and $\Delta_{\text{sat}}$ represent the maximum position errors for the nodes and satellites, respectively. The estimated positions are obtained by adding random errors within these bounds to the true positions.

For the $n$-th node with true position ($x_{\text{node},n}$, $z_{\text{node},n}$) in the $k$-th node sub-region, the estimated position is

\begin{equation}
\begin{aligned}
x_{\text{node},n}^{\text{est}} &= x_{\text{node},n} + \delta x_{\text{node},n}, \\
z_{\text{node},n}^{\text{est}} &= z_{\text{node},n} + \delta z_{\text{node},n},
\end{aligned}
\end{equation}
where $\delta x_{\text{node},n}$ and $\delta z_{\text{node},n}$ are random variables uniformly distributed in $[-\Delta_{\text{node,GPS}}, \Delta_{\text{node,GPS}}]$.

Similarly, the estimated position of the $k$-th satellite with true position ($x_{\text{satB},k}$, $z_{\text{satB},k}$) is

\begin{equation}
\begin{aligned}
x_{\text{satB},k}^{\text{est}} &= x_{\text{satB},k} + \delta x_{\text{sat},k}, \\
z_{\text{satB},k}^{\text{est}} &= z_{\text{satB},k} + \delta z_{\text{sat},k},
\end{aligned}
\end{equation}
where $\delta x_{\text{sat},k}$ and $\delta z_{\text{sat},k}$ are random variables uniformly distributed in $[-\Delta_{\text{sat}}, \Delta_{\text{sat}}]$. In practice, the satellite position can be estimated with an accuracy better than 1 centimeter~\cite{rudenko2023radial}. Therefore, we can assume that the true satellite position is effectively known and set $\Delta_{\text{sat}}$ to zero.

Using the estimated positions, we calculate the estimated distance between each node and its corresponding satellite

\begin{equation}
d_{\text{satB-node},n}^{\text{est}} = \sqrt{\left( x_{\text{satB},k}^{\text{est}} - x_{\text{node},n}^{\text{est}} \right)^2 + \left( z_{\text{satB},k}^{\text{est}} - z_{\text{node},n}^{\text{est}} \right)^2}.
\end{equation}

We also calculate the estimated radial distance of the node from Earth's center
\begin{equation}
r_{\text{node},n}^{\text{est}} = \sqrt{\left( x_{\text{node},n}^{\text{est}} \right)^2 + \left( z_{\text{node},n}^{\text{est}} \right)^2}.
\end{equation}

The estimated angular positions of the node and satellite are

\begin{equation}
\begin{aligned}
\theta_{\text{node},n}^{\text{est}} &= \arctan\left( x_{\text{node},n}^{\text{est}}, z_{\text{node},n}^{\text{est}} \right), \\
\theta_{\text{satB},k}^{\text{est}} &= \arctan\left( x_{\text{satB},k}^{\text{est}}, z_{\text{satB},k}^{\text{est}} \right).
\end{aligned}
\end{equation}

The estimated angle between the node and its corresponding satellite is
\begin{equation}
\Delta\theta_{\text{satB-node},n}^{\text{est}} = \theta_{\text{satB},k}^{\text{est}} - \theta_{\text{node},n}^{\text{est}}.
\end{equation}

We compute an intermediate angle $\alpha_n$ using the Law of Sines
\begin{equation}
\alpha_n = \arcsin\left( \frac{r_{\text{node},n}^{\text{est}}}{d_{\text{satB-node},n}^{\text{est}}} \sin\Delta\theta_{\text{satB-node},n}^{\text{est}} \right).
\end{equation}

The estimated angle between the satellite and the steering direction is
\begin{equation}
\Delta\theta_{\text{satB-steer},k}^{\text{est}} = \theta_{\text{steer}} - \theta_{\text{satB},k}^{\text{est}}.
\end{equation}

The angle between the node-satellite line and the steering direction is then
\begin{equation}
\beta_n = \alpha_n - \Delta\theta_{\text{satB-steer},k}^{\text{est}}.
\end{equation}

The excess path length due to the angle $\beta_n$ is calculated by
\begin{equation}
\delta d_{\text{GPS},n} = d_{\text{satB-node},n}^{\text{est}} \left( 1 - \cos \beta_n \right).
\end{equation}

The GPS correction phase is then
\begin{equation}
\phi_{\text{GPS},n} = \frac{2\pi}{\lambda} \delta d_{\text{GPS},n}.
\end{equation}

The corrected calibration phase for each node is
\begin{equation}
\phi_{\text{cal,GPS},n} = \phi_{\text{cal,satB},n} - \phi_{\text{GPS},n},
\end{equation}
where $\phi_{\text{cal,satB},n}$ is the original calibration phase without GPS correction, given by
\begin{equation}
\phi_{\text{cal,satB},n} = \frac{2\pi}{\lambda} d_{\text{satB-node},n} + \phi_{\text{satB-steer},k}.
\end{equation}

The array factor after incorporating GPS-based correction is
\begin{equation}
AF_{\text{satB,GPS}} = \sum_{k=1}^{N_{\text{satB}}} \sum_{n \in \mathcal{R}_k} \exp\left( j \phi_{\text{geom},n} + j \phi_{\text{cal,GPS},n} \right).
\end{equation}

The array gain with GPS correction is then
\begin{equation}
G_{\text{satB,GPS}} = 10 \log_{10}  \left| AF_{\text{satB,GPS}} \right|^2 .
\end{equation}

In summary, the finite distance of Satellite B causes the incoming wavefront to have curvature, leading to phase discrepancies among the nodes when compared to plane wavefronts from astronomical sources at infinite distances. By using GPS estimates of the node and satellite positions, we calculate the angle between the node-satellite line and the steering direction, $\beta_n$. This angle results in an excess path length $\delta d_{\text{GPS},n}$, which introduces phase errors in the calibration process.

By computing $\delta d_{\text{GPS},n}$ and the corresponding correction phase $\phi_{\text{GPS},n}$, we adjust the calibration phase for each node to account for these errors, effectively aligning the phase of the received signals as if they were from an astronomical source at infinite distance.

\subsection{Analysis Results for Calibration Using Satellite A}

Fig.~\ref{fig:satA_diff_baseline} illustrates the normalized array gain for Satellite A over different baseline lengths, with the satellite height $h_{\rm{satA}}$ varied among $10^{12}$\,meters, $10^{13}$\,meters, and $10^{14}$\,meters. The figure shows that the array gain decreases as the baseline length $L_{\rm{baseline}}$ increases. For a given baseline length, increasing the satellite height improves the array gain. This improvement is attributed to the reduction in wavefront curvature effects at higher altitudes, leading to more uniform phase fronts across the array. However, when the baseline length reaches $12 \times 10^{6}$\,meters—approximately the diameter of the Earth—the required satellite height $h_{\rm{satA}}$ exceeds $10^{14}$\,meters to achieve acceptable array gain. Such altitudes are impractically large for real-world applications, highlighting the limitations of using a single calibration satellite for extensive baseline lengths.

\begin{figure}[htp!]
      \centering
      \includegraphics[scale=\plotsize]{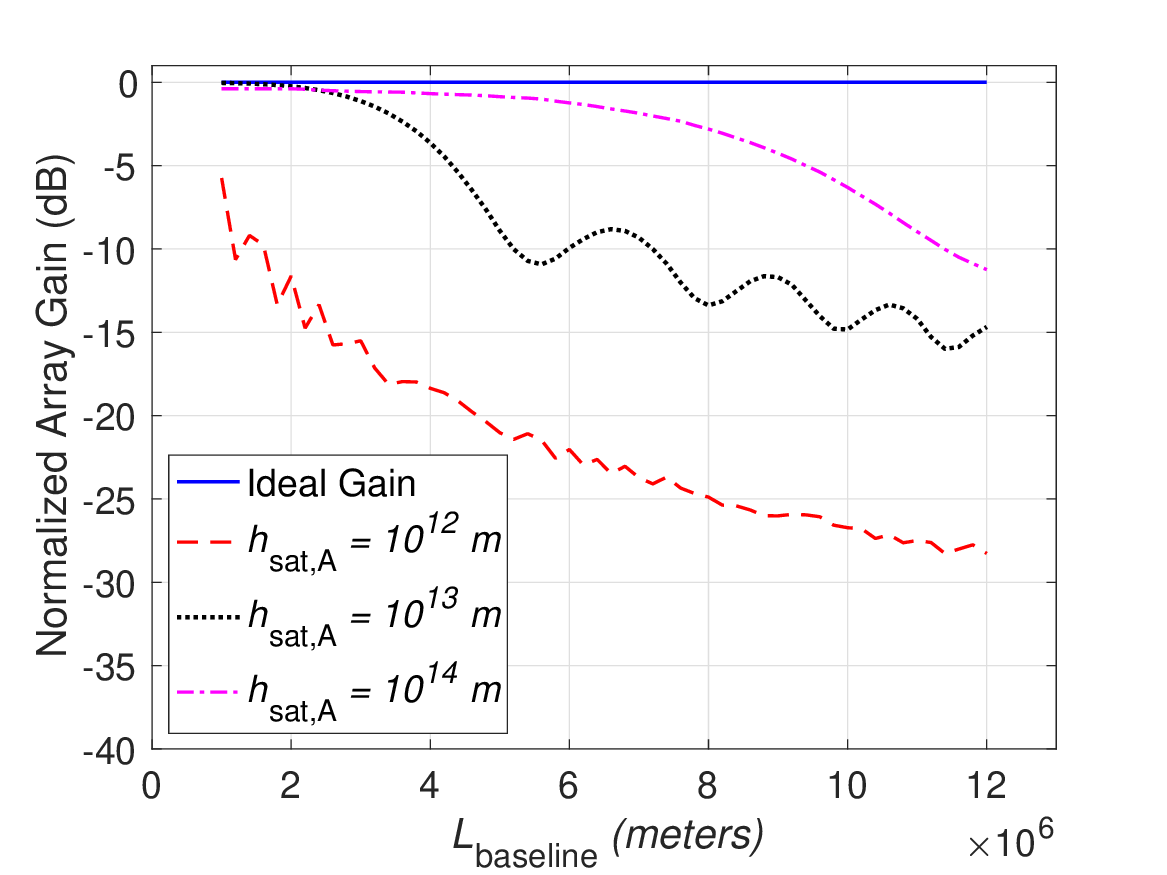}
      \caption{Normalized array gain for Satellite A over different baseline lengths. The satellite heights are $10^{12}$\,m, $10^{13}$\,m, and $10^{14}$\,m.}
      \label{fig:satA_diff_baseline}
\end{figure}

\subsection{Analysis Results for Calibration Using Satellite B}

To address the limitations of the single-satellite calibration approach, we present analysis results for the proposed phase calibration method using Satellite B. The analysis explores various parameters such as node localization estimation error, steering angles, number of satellites, satellite heights, and baseline lengths.

\begin{figure}[htp!]
      \centering
      \includegraphics[scale=\plotsize]{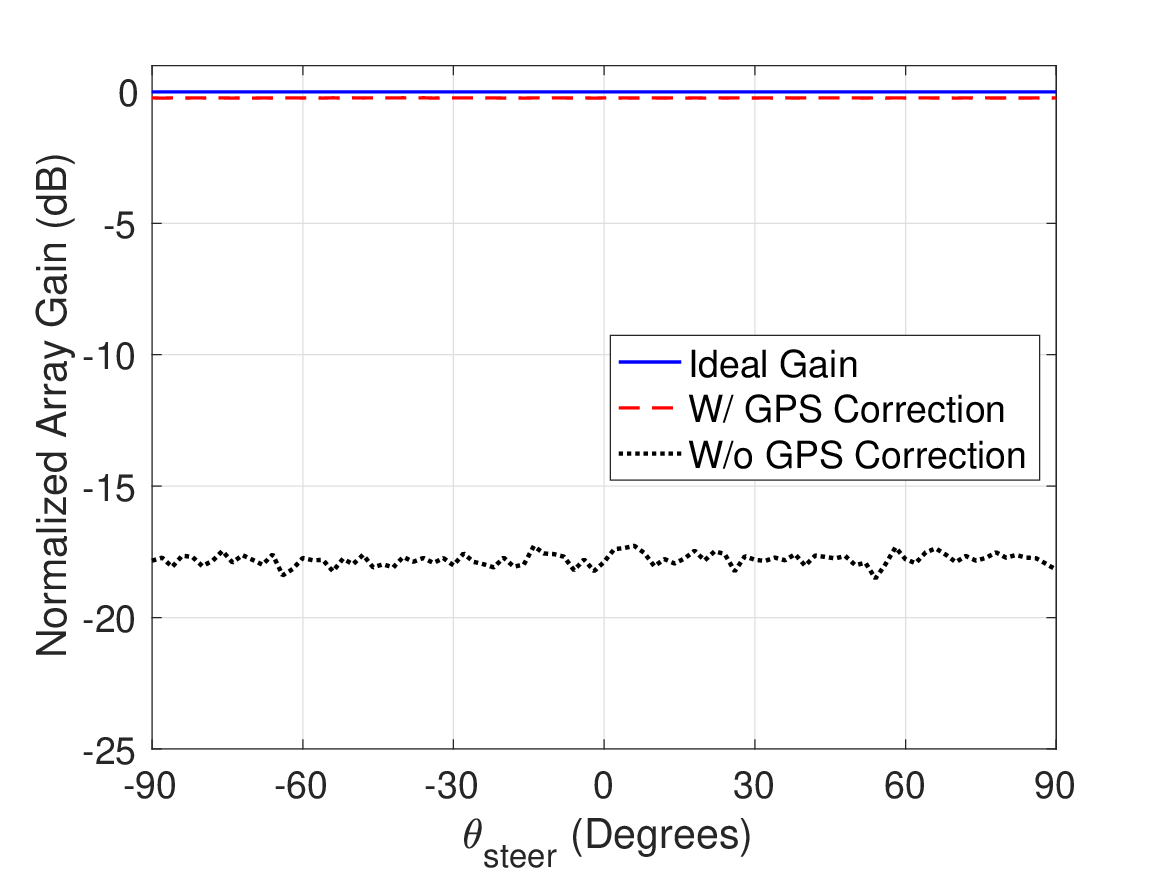}
      \caption{Normalized array gain over different steering angles, with and without GPS correction.}
      \label{fig:satB_GPS_correction}
\end{figure}

\subsubsection{Effectiveness of GPS Correction}

Fig.~\ref{fig:satB_GPS_correction} illustrates the normalized array gain over different steering angles, with and without GPS correction. The node localization estimation error with GPS $\Delta_{\rm{node,GPS}}$ is set to 5 meters, the height of Satellite B is $h_{\rm{satB}} = 5\times10^{6}$ meters, the baseline length is $L_{\rm{baseline}} = 12,\!000$ kilometers, and the number of satellites is $N_{\rm{satB}} = 1,\!000$.

The analysis results clearly demonstrate that GPS correction significantly enhances the array gain across all steering angles, achieving performance nearly equivalent to the ideal array gain $G_{\rm{ideal}}$ as defined in equation (\ref{eq:ideal_gain}). In the absence of GPS correction, the array gain experiences a substantial degradation of over 18 dB, primarily due to phase errors introduced by the wavefront curvature of Satellite B. GPS correction effectively mitigates these phase discrepancies by providing accurate position estimates, which ensures better alignment of the received signal phases and thereby restores the array gain to levels close to the ideal array gain.

\subsubsection{Impact of GPS Estimation Error}

Fig.~\ref{fig:satB_diff_GPSError} illustrates the normalized array gain for Satellite B as a function of GPS estimation error $\Delta_{\rm{node,GPS}}$, with the number of satellites $N_{\text{satB}}$ varied among 20, 100, and 1,000. The height of Satellite B is $h_{\rm{satB}} = 5\times10^{6}$ meters, the baseline length is $L_{\rm{baseline}} = 12,\!000$ kilometers, and the steering angle is $\theta_{\rm{steer}} = 0^{\circ}$.

\begin{figure}[htp!]
      \centering
      \includegraphics[scale=\plotsize]{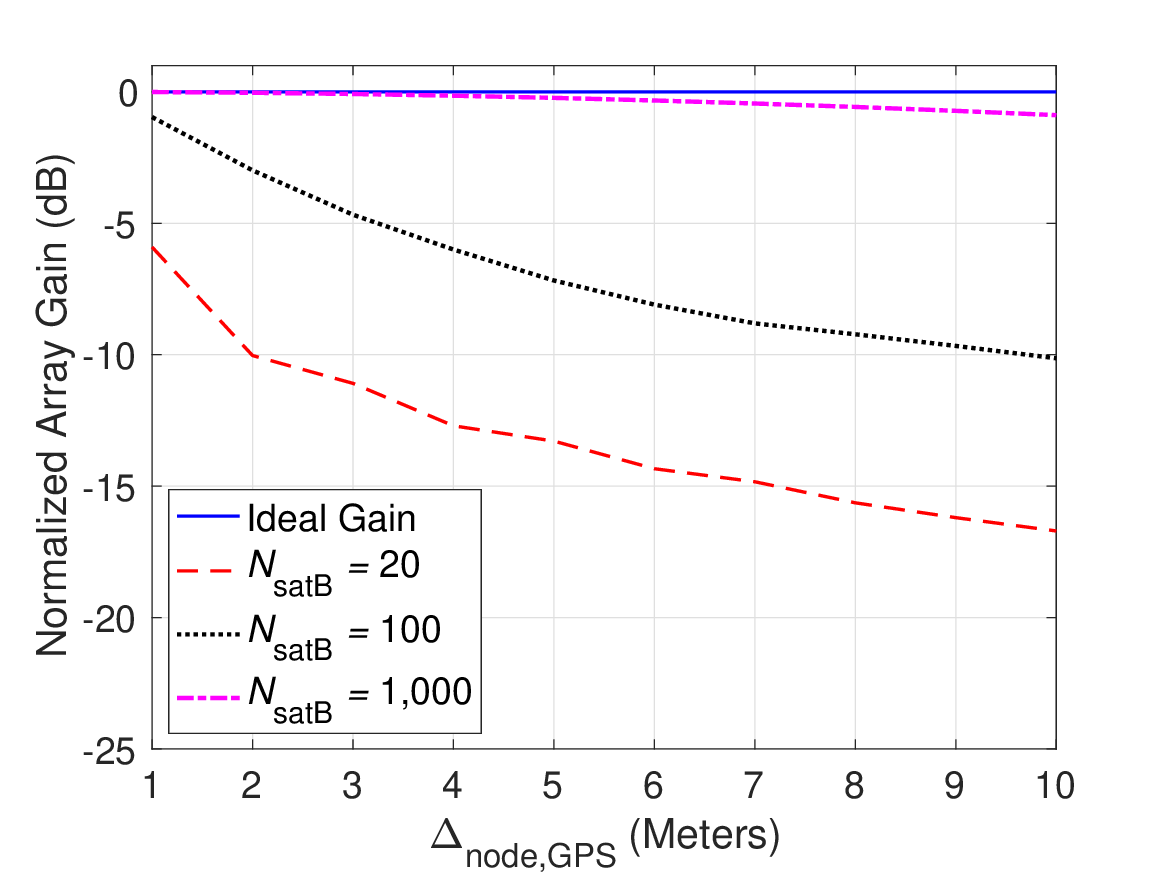}
      \caption{Normalized array gain with different GPS estimation errors $\Delta_{\rm{node,GPS}}$.}
      \label{fig:satB_diff_GPSError}
\end{figure}

The results clearly show that the array gain decreases as the GPS estimation error increases. This degradation occurs because larger estimation errors in the node positions lead to less effective correction of the curvature of the satellite calibration signal's wavefront. As a result, the phases of the received signals become misaligned, leading to less coherent combining and a reduction in array gain.

For a given GPS estimation error $\Delta_{\rm{node,GPS}}$, increasing the number of satellites $N_{\text{satB}}$ enhances the array gain. This improvement can be explained by the fact that increasing the number of satellites divides the array into more node sub-regions, each calibrated by its own satellite. This division effectively reduces the baseline length within each node sub-region. As observed from the results in Fig.~\ref{fig:satA_diff_baseline}, a smaller baseline length corresponds to a larger array gain due to reduced phase errors over shorter distances. Therefore, increasing $N_{\text{satB}}$ mitigates the adverse effects of GPS estimation errors by limiting phase discrepancies within smaller sub-arrays. This enhancement is especially noticeable when the GPS estimation error is relatively large, as the impact of position inaccuracies is confined within smaller regions, allowing for more effective phase alignment and improved overall array performance.

\subsubsection{Impact of Beam Steering Angles}

Fig.~\ref{fig:satB_different_steerAngle} presents the normalized array gain for Satellite B as a function of steering angle $\theta_{\text{steer}}$, with the number of satellites $N_{\text{satB}}$ varied among 20, 50, and 500. In this analysis, the node localization estimation error with GPS is set to $\Delta_{\rm{node,GPS}} = 5$ meters, the height of Satellite B is $h_{\rm{satB}} = 5 \times 10^{6}$ meters, and the baseline length is $L_{\rm{baseline}} = 12,\!000$ kilometers.

\begin{figure}[htp!]
      \centering
      \includegraphics[scale=\plotsize]{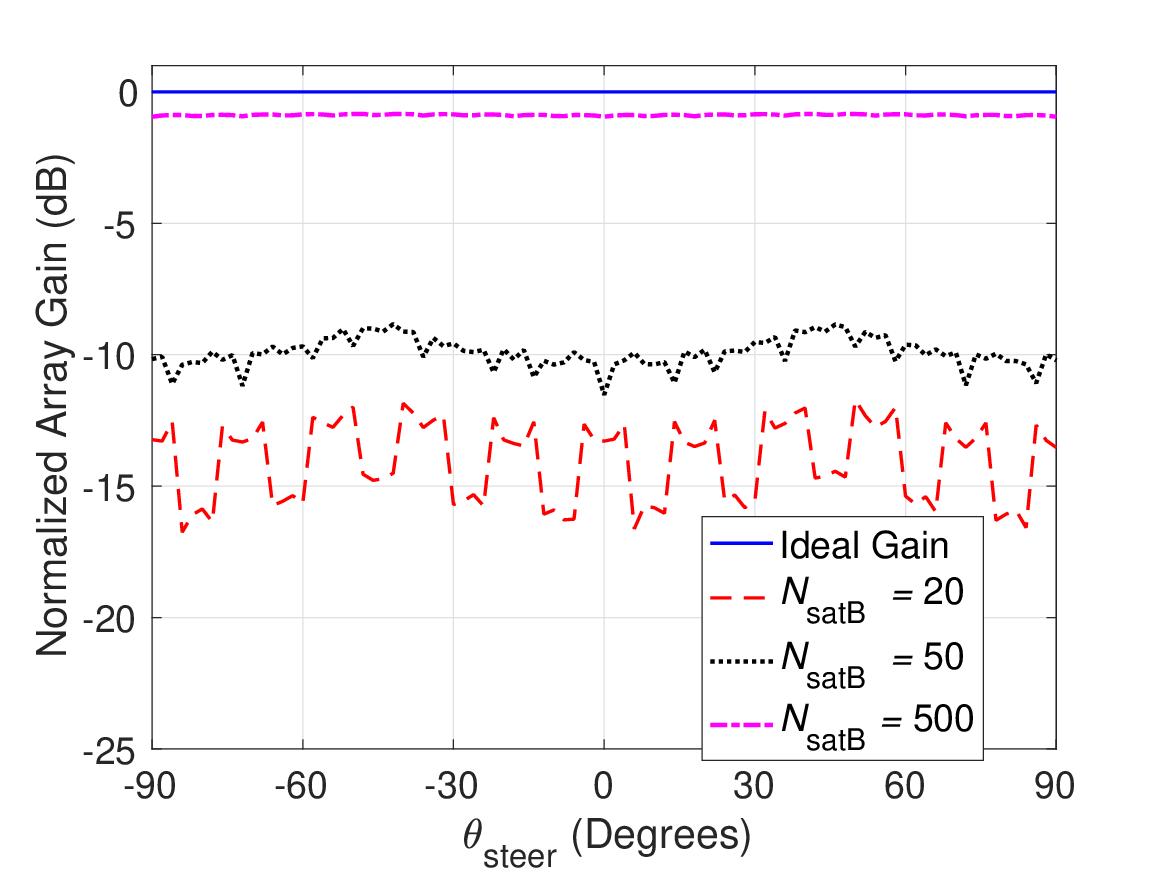}
      \caption{Normalized array gain with different steering angles for various numbers of satellites.}
      \label{fig:satB_different_steerAngle}
\end{figure}

When $N_{\text{satB}} = 20$, the array gain is over 13 dB less than the ideal array gain at certain steering angles. The dips in the array gain as $\theta_{\text{steer}}$ varies are due to significant phase errors when the beam is steered between the coverage areas of adjacent satellites.

As the number of satellites increases to $N_{\text{satB}} = 50$, the ripple effect is mitigated, and the array gain improves across the steering angle range. This is because the additional satellites provide better coverage and reduce the gaps between calibration regions, resulting in more consistent scanned beams.

When $N_{\text{satB}}$ is further increased to 500, the array gain becomes almost identical to the ideal array gain and remains nearly constant across all steering angles $\theta_{\text{steer}}$. The high density of satellites ensures continuous scanned beams, effectively eliminating phase errors due to satellite transitions.

These results indicate that by sufficiently increasing the number of calibration satellites, the proposed telescope system can achieve a nearly isotropic gain response. Moreover, theoretically, an infinite number of beams with uniform gain over the full scan range can be implemented simultaneously using digital beamforming techniques. This represents a significant improvement over traditional radio telescopes, which are typically limited to only up to tens of simultaneous beams~\cite{roshi2018performance}.

\subsubsection{Impact of Number of Satellites}

Fig.~\ref{fig:satB_different_NsatB} depicts the normalized array gain for Satellite B with different numbers of satellites $N_{\text{satB}}$. The satellite heights $h_{\text{satB}}$ are $1\times10^{6}$\,m, $5\times10^{6}$\,m, and $2\times10^{7}$\,m. The node localization estimation error with GPS is set to $\Delta_{\rm{node,GPS}} = 5$ meters, the baseline length is $L_{\rm{baseline}} = 12,\!000$ kilometers, and the steering angle is $\theta_{\rm{steer}} = 0^{\circ}$.

\begin{figure}[htp!]
      \centering
      \includegraphics[scale=\plotsize]{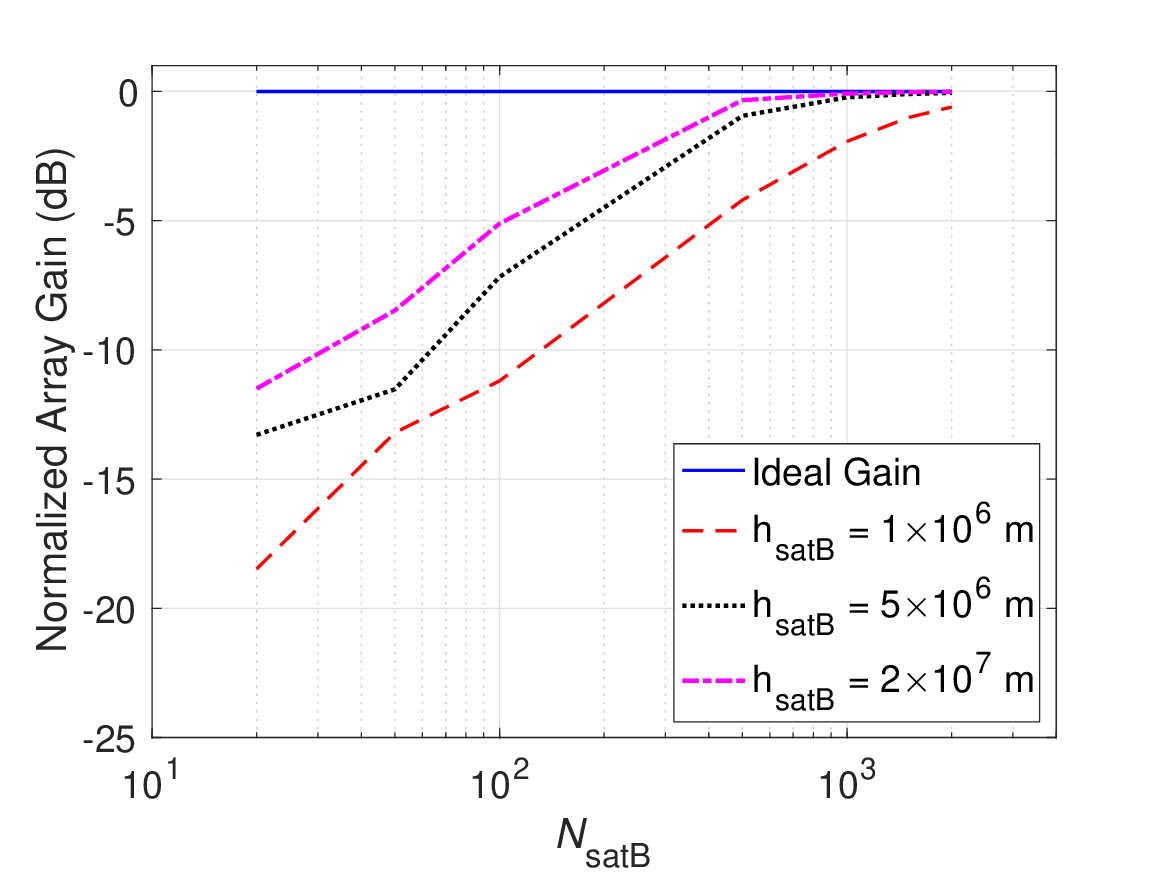}
      \caption{Normalized array gain with different numbers of satellites at various satellite heights.}
      \label{fig:satB_different_NsatB}
\end{figure}

The results indicate that for a given satellite height, increasing the number of satellites enhances the array gain. Moreover, higher satellite altitudes result in better performance due to reduced wavefront curvature effects. This demonstrates that both the number of satellites and their altitude play crucial roles in achieving optimal array gain.

\subsubsection{Impact of Satellite Height}

Fig.~\ref{fig:satB_different_height} illustrates the normalized array gain for Satellite B over different satellite heights $h_{\text{satB}}$, with the number of satellites $N_{\text{satB}}$ varied among 20, 100, and 500. The node localization estimation error with GPS is set to $\Delta_{\rm{node,GPS}} = 5$ meters, the baseline length is $L_{\rm{baseline}} = 12,\!000$ kilometers, and the steering angle is $\theta_{\rm{steer}} = 0^{\circ}$.

The array gain improves with increasing satellite height due to the flattening of the wavefronts at higher altitudes. This effect is more pronounced when combined with a higher number of satellites, which further reduces phase errors due to the smaller baselines of the node sub-regions.

\begin{figure}[htp!]
      \centering
      \includegraphics[scale=\plotsize]{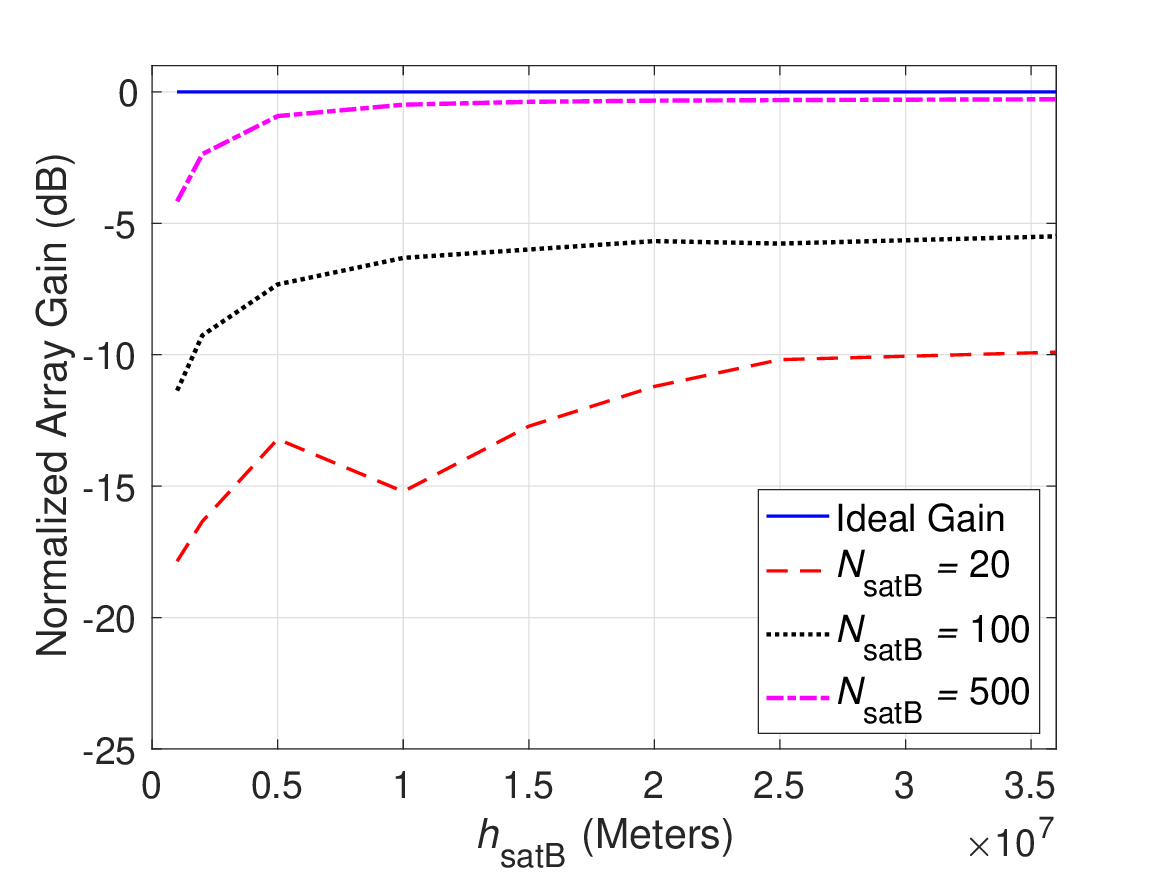}
      \caption{Normalized array gain over different satellite heights for various numbers of satellites.}
      \label{fig:satB_different_height}
\end{figure}

\subsubsection{Impact of Baseline Length}

Fig.~\ref{fig:satB_diff_baseline} shows the normalized array gain for Satellite B over different baseline lengths $L_{\rm{baseline}}$, with the number of satellites $N_{\text{satB}}$ set to 50, 100, and 500. The node localization estimation error with GPS is set to $\Delta_{\rm{node,GPS}} = 5$ meters, the satellite height is $h_{\rm{satB}} = 5 \times 10^{6}$ meters, and the steering angle is $\theta_{\rm{steer}} = 0^{\circ}$. The results indicate that as the baseline length increases, the array gain increased due to increased number of nodes. Increasing the number of satellites helps mitigate this degradation from the ideal gain by providing more precise phase calibration over the extended baseline.

\begin{figure}[htp!]
      \centering
      \includegraphics[scale=\plotsize]{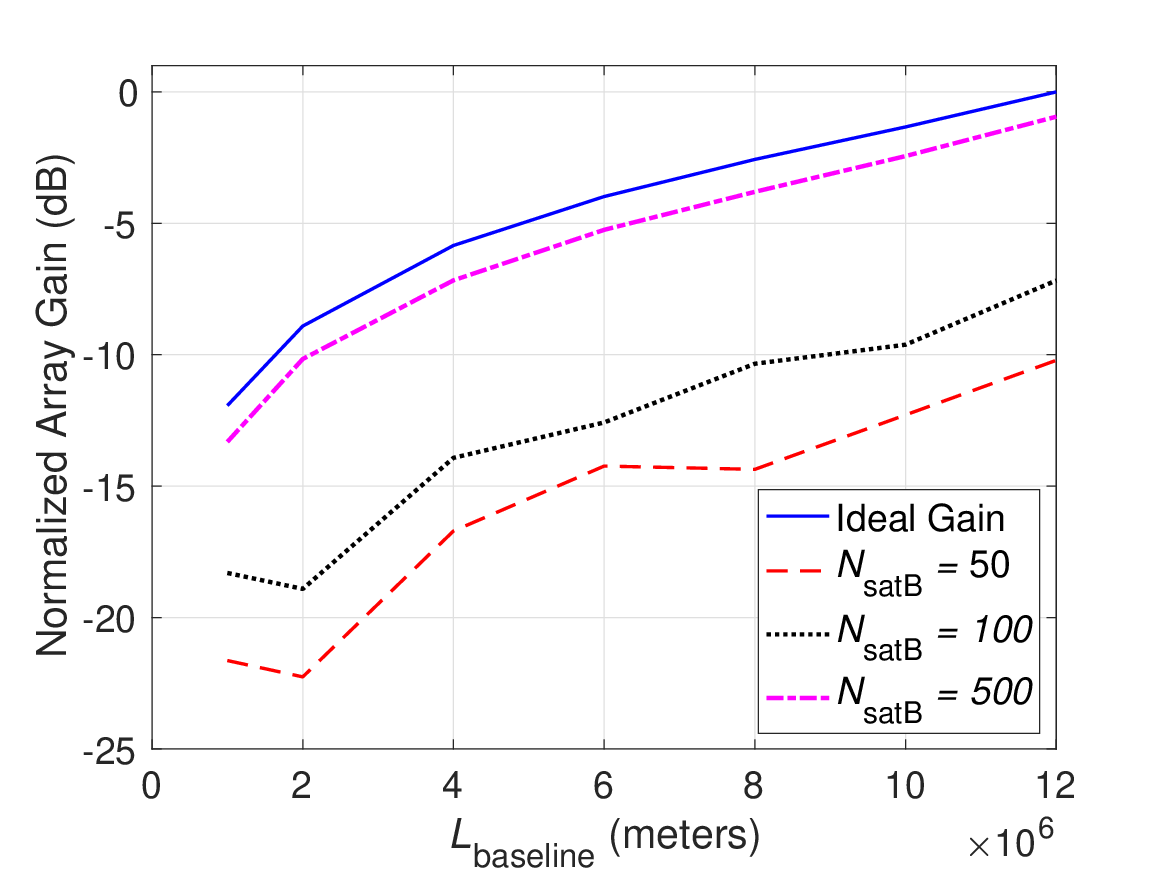}
      \caption{Normalized array gain over different baseline lengths for various numbers of satellites.}
      \label{fig:satB_diff_baseline}
\end{figure}

\subsubsection{Required Number of Satellites for Full-Spherical 3-D Field-of-View Coverage}

\textcolor{black}{The satellite requirements derived from the 2-D cut-plane analysis can be extended to full 3-D spherical field-of-view coverage using a geometric scaling argument. In the 2-D case, uniform angular coverage is achieved by distributing $N_{\mathrm{satB}}$ calibration satellites across the angular plane, resulting in an approximate angular separation
\begin{equation}
\theta \approx \frac{2\pi}{N_{\mathrm{satB}}}.
\end{equation}}

\textcolor{black}{For full-spherical coverage, calibration signals must uniformly span the entire solid angle of $4\pi$ steradians with comparable angular resolution in both azimuth and elevation. Assuming an angular resolution on the order of $\theta$ in each dimension, the required number of satellite directions scales as
\begin{equation}
N_{\mathrm{sat}}^{(3\mathrm{D})} \approx \frac{4\pi}{\theta^2}
= \frac{4\pi}{\left(2\pi/N_{\mathrm{sat}}\right)^2}
= \frac{N_{\mathrm{sat}}^2}{\pi}.
\end{equation}
This relationship indicates that the total number of calibration satellites required for full-spherical coverage increases approximately with the square of the 2D requirement, reflecting the need to resolve angular variation over both elevation and azimuth.}

\textcolor{black}{Table~\ref{tab:sat_scaling} summarizes the estimated number of calibration satellites required to achieve full-spherical field-of-view coverage under different GPS positioning accuracies and orbital regimes, including low Earth orbit (LEO) at 2,000 km, medium Earth orbit (MEO) at 10,000 km, and geostationary Earth orbit (GEO) at 36,000 km. The results highlight the strong dependence of the required satellite count on localization precision, and indicate that continued improvements in GPS accuracy can substantially reduce the calibration infrastructure needed for global IoT-enabled radio telescope operation.}

\begin{table}[t]
\centering
\caption{Estimated number of calibration satellites required for full-spherical field-of-view coverage.}
\label{tab:sat_scaling}
\footnotesize
\setlength{\tabcolsep}{3pt}
\renewcommand{\arraystretch}{1.15}
\begin{tabular}{c c c c}
\hline
\thead{GPS Positional Accuracy\\($\Delta_{\rm{node,GPS}}$)} &
\thead{LEO Satellite\\Count} &
\thead{MEO Satellite\\Count} &
\thead{GEO Satellite\\Count} \\
\hline
0.5 meters & 7,000   & 800  & 300  \\
1 meter   & 29,000  & 32,000 & 16,000 \\
3 meters   & 200,000 & 29,000  & 13,000  \\
\hline
\end{tabular}
\end{table}

\section{System Signal Model}
\label{section:system_signal_model}
In this section, we present a comprehensive signal model to better understand the proposed IoT-enabled telescope. The model captures the entire signal chain of each IoT node, from reception to final beamformed output, while incorporating hardware and system noise variability, local oscillator (LO) offset, interference mitigation, time-division calibration/observation scheduling, calibration methods, and output SNR calculations.

\subsection{Node Distribution and Hardware Variability}
We consider a one-dimensional array of $N$ IoT nodes randomly distributed along the $x$-axis over an aperture length $L$. The position of the $n$-th node is given by
\begin{equation}
    x_n = L \, (u_n - 0.5),
\end{equation}
\textcolor{black}{where $u_n \sim \mathcal{U}[0,1]$ is a uniform random variable to simulate uneven spatial deployment of the global nodes.}

Each node is equipped with an antenna and a receiver front-end whose performance is characterized by variability in antenna gain and noise figure. In particular, we model the antenna gain (in dB) as
\begin{equation} \label{eq:Gn}
    G_{n,\mathrm{dB}} = G_{\mathrm{ant}} + \sigma_G \, z^G_n,
\end{equation}
and the noise figure (in dB) as
\begin{equation} \label{eq:NF_n}
    \mathrm{NF}_{n,\mathrm{dB}} = \mathrm{NF}_0 + \sigma_{\mathrm{NF}} \, z^{\mathrm{NF}}_n,
\end{equation}
where $G_{\mathrm{ant}}$ and $\mathrm{NF}_0$ are the nominal antenna gain and noise figure, $\sigma_G$ and $\sigma_{\mathrm{NF}}$ are their standard deviations, and $z^G_n$, $z^{\mathrm{NF}}_n \sim \mathcal{N}(0,1)$ are independent standard normal random variables. Converting to linear scale yields
\begin{equation}
    G_n = 10^{G_{n,\mathrm{dB}}/10}, \quad \mathrm{NF}_n = 10^{\mathrm{NF}_{n,\mathrm{dB}}/10}.
\end{equation}
\textcolor{black}{The variations in gain and noise figure across nodes reflects hardware diversity and potential indoor environments.}

The effective antenna area at node $n$ is computed as
\begin{equation}
    A_{\mathrm{eff},n} = \frac{G_n \, \lambda_c^2}{4\pi},
\end{equation}
where $\lambda_c = c/f_c$ is the wavelength at the center frequency $f_c$.

Each node's receiver noise temperature is modeled by
\begin{equation}
    T_{\mathrm{noise},n} = T_0 \cdot \mathrm{NF}_n,
\end{equation}
with $T_0=290$~K. The corresponding noise power over a bandwidth $B$ is
\begin{equation}
    P_{\mathrm{noise},n} = k_B \, T_{\mathrm{noise},n} \, B,
\end{equation}
where $k_B$ is the Boltzmann constant.

\subsection{Signal Modeling}
During the calibration-on period, the calibration signal received at node $n$ is modeled as
\begin{equation} \label{eq:s_cal}
    s^{\mathrm{cal\mbox{-}on}}_n(t) = A^{\mathrm{cal}}_n \cos\Big(2\pi f_c t - k_c\,x_n\cos\theta_{\mathrm{cal}}\Big),
\end{equation}
where
\begin{equation}
    A^{\mathrm{cal}}_n = \sqrt{2\,P_{\mathrm{cal}}\,A_{\mathrm{eff},n}},
\end{equation}
$P_{\mathrm{cal}}$ is the calibration signal power density, $k_c = \frac{2\pi}{\lambda_c}$, and $\theta_{\mathrm{cal}}$ is the known direction of the calibration signal.

During the observation period, the astronomical signal is given by
\begin{equation} \label{eq:s_obs}
    s^{\mathrm{obs}}_n(t) = A^{\mathrm{obs}}_n \cos\Big(2\pi f_c t - k_c\,x_n\cos\theta_{s}\Big),
\end{equation}
with
\begin{equation}
    A^{\mathrm{obs}}_n = \sqrt{2\,P_{\mathrm{astro}}\,A_{\mathrm{eff},n}},
\end{equation}
where $P_{\mathrm{astro}}$ is the astronomical signal power density and $\theta_{s}$ is its direction of arrival.

\subsection{Noise and Interference}
The received signal at node $n$ is subject to additive noise $n_n(t)$, modeled as a zero-mean Gaussian process with variance $\sigma_n^2 = P_{\mathrm{noise},n}$:
\begin{equation}
    n_n(t) \sim \mathcal{N}\big(0,\; \sigma_n^2\big).
\end{equation}

Furthermore, interference from external sources is incorporated into the model. Assuming there are $Q$ interferers, the interference at node $n$ is modeled as
\begin{equation}
    i_n(t) = \sum_{q=1}^{Q} A^{\mathrm{int}}_{q,n} \cos\Big(2\pi f_{q} t - k_c\,x_n\cos\theta_{q} + \varphi_q\Big),
\end{equation}
where the amplitude of the $q$-th interferer at node $n$ is given by
\begin{equation}
    A^{\mathrm{int}}_{q,n} = \sqrt{2\,P^{\mathrm{int}}_{q}\,A_{\mathrm{eff},n}}.
\end{equation}
Here, $P^{\mathrm{int}}_{q}$ denotes the interference power density of the $q$-th interferer, $f_{q}$ is its carrier frequency, $\theta_{q}$ is its direction of arrival, and $\varphi_q$ is an arbitrary phase. \textcolor{black}{The arbitrary phase terms $\varphi_q$ is used to emulate time-varying communication interference with randomized spatial and spectral characteristics.}

Thus, the total RF signal received at node $n$ is
\begin{equation} \label{eq:r_n_rf}
    r_n(t) = s_n(t) + i_n(t) + n_n(t),
\end{equation}
where
\[
s_n(t) = 
\begin{cases}
    s^{\mathrm{cal\mbox{-}on}}_n(t), & \text{during calibration-on},\\[1mm]
    s^{\mathrm{obs}}_n(t), & \text{during observation}.
\end{cases}
\]
In other words, each node receives its desired signal $s_n(t)$ (either the calibration signal or the astronomical signal) superimposed with the aggregate interference from all $Q$ external sources and additive noise.

\subsection{Successive Interference Cancellation and Beamforming Suppression}
\label{subsec:sic}

In the proposed IoT-enabled telescope, individual nodes are expected to operate in spectrum-sharing environments where strong communication signals may coexist with extremely weak astronomical signals. If left untreated, such interference can dominate the received waveform and reduce the effectiveness of subsequent distributed beamforming and correlation. SIC is therefore introduced as a front-end mitigation step to suppress dominant interferers and, more importantly, to decorrelate residual interference across spatially distributed nodes. This ensures that interference does not combine coherently in the beamforming stage, allowing the astronomical signal to be enhanced through spatial coherence.

Starting from the received RF signal in \eqref{eq:r_n_rf}, each node estimates and subtracts the aggregate interference component. The resulting post-SIC signal at node $n$ is modeled as
\begin{equation}
\tilde{r}_n(t) = r_n(t)-\hat{i}_n(t)
= s_n(t)+n_n(t)+e_n(t),
\end{equation}
where $e_n(t)=i_n(t)-\hat{i}_n(t)$ denotes the residual interference after SIC. Distributed beamforming toward the astronomical direction $\theta_s$ is then applied using steering weights $w_n(\theta_s)$, yielding the beamformer output
\begin{equation}
y(t)=\frac{1}{N}\sum_{n=1}^{N} w_n^*(\theta_s)\,\tilde{r}_n(t)
= y_s(t)+y_w(t)+y_e(t),
\end{equation}
where $y_e(t)=\frac{1}{N}\sum_{n=1}^{N} w_n^*(\theta_s)e_n(t)$ represents the contribution of the residual interference.

Unlike the desired astronomical signal, which is phase-aligned across nodes after steering, the residual interference is generally not spatially coherent. It is therefore reasonable to model
\begin{equation}
\mathrm{E}\!\left[e_n(t)e_m^*(t)\right]\approx 0,\qquad n\neq m,
\end{equation}
which implies that the residual interference combines incoherently across the array. Under this assumption, the beamformed residual interference power satisfies
\begin{equation}
\mathrm{E}\!\left[|y_e(t)|^2\right]
=\frac{1}{N^2}\sum_{n=1}^{N}\mathrm{E}\!\left[|e_n(t)|^2\right]
\propto \frac{1}{N}.
\end{equation}
As a result, even when SIC is imperfect at individual nodes, the combination of SIC and distributed beamforming effectively suppresses interference as the number of IoT nodes increases, while the desired astronomical signal combines coherently.

\subsection{LO Offset and Baseband Conversion}
Each node downconverts the RF signal to baseband using its LO. Although ideally $f_{LO}=f_c$, each node experiences a small frequency offset $\Delta f_n$ and an initial phase offset $\phi_n$, so that
\begin{equation}
    f_{LO,n} = f_c + \Delta f_n.
\end{equation}
The mixing operation is performed by multiplying the clipped signal by the LO complex exponential:
\begin{equation}
    m_n(t) = e^{-j\,(2\pi f_{LO,n} t + \phi_n)}.
\end{equation}
After mixing and ideal low-pass filtering, the complex baseband signal is obtained. For the calibration signal, this yields
\begin{equation} \label{eq:y_cal_base}
    y^{\mathrm{cal\mbox{-}on}}_n(t) = \frac{A^{\mathrm{cal}}_n}{2}\, e^{-j\,(k_c\,x_n\cos\theta_{\mathrm{cal}} + \phi_n + 2\pi \Delta f_n t)}  + \tilde{n}^{\mathrm{cal}}_n(t),
\end{equation}
where $\tilde{n}^{\mathrm{cal}}_n(t)$ represents the filtered noise. Here, we assume that the calibration signal can be distinguished from interference by utilizing its known code sequence. Similarly, for the astronomical signal during observation, we have
\begin{equation} \label{eq:y_obs_base}
    y^{\mathrm{obs}}_n(t) = \frac{A^{\mathrm{obs}}_n}{2}\, e^{-j\,(k_c\,x_n\cos\theta_{s} + \phi_n + 2\pi \Delta f_n t)} + \tilde{n}^{\mathrm{obs}}_n(t) + \tilde{i}_n(t),
\end{equation}
with $\tilde{n}^{\mathrm{obs}}_n(t)$ and $\tilde{i}_n(t)$ being the baseband noise and interference components, respectively. In practice, the real downconverted signal is converted to its analytic form using the Hilbert transform, yielding the equivalent complex baseband signals.

\subsection{Time-Division for Calibration and Observation Modes}
The system alternates between two modes:

\begin{enumerate}
    \item \textit{Calibration Mode:} Over a calibration cycle of duration $T_{\mathrm{cal}}$, a transmitter emits a known signal for a period $T_{\mathrm{cal\mbox{-}on}}$, with the remaining time
    \[
    T_{\mathrm{cal\mbox{-}off}} = T_{\mathrm{cal}} - T_{\mathrm{cal\mbox{-}on}}
    \]
    used to measure noise. During the calibration-on period, the received signals at each node are employed to estimate phase and amplitude differences across the array.
    
    \item \textit{Observation Mode:} After calibration, an observation period of duration $T_{\mathrm{obs}}$ is used to acquire astronomical data. During this period, each node receives the weak astronomical signal in addition to noise and interference. ADC clipping limits strong interference, and the downconverted baseband signals are combined using the calibration-derived beamforming weights.
\end{enumerate}

This time-division scheme prevents strong calibration signals from contaminating astronomical observations while enabling accurate noise measurement and calibration.

\subsection{Calibration Methods}
To enable coherent combination across the array, calibration is performed using the calibration signal. Two methods are considered:

\subsubsection{Phase Alignment Calibration (PAC)}
In PAC, the relative phase differences between nodes are estimated via cross-correlation during the calibration-on period. Let
\begin{equation}
    y_n^{\mathrm{cal\mbox{-}on}}(t) \approx \frac{A^{\mathrm{cal}}_n}{2}\, e^{-j\,[k_c\,x_n\cos\theta_{\mathrm{cal}} + \phi_n]} + \tilde{n}_n^{\mathrm{cal\mbox{-}on}}(t).
\end{equation}
The phase difference between node $n$ and a reference node (say, node 1) is estimated as
\begin{equation}
    \hat{\Delta\phi}_n = \arg\!\left\{ \frac{1}{T_{\mathrm{cal\mbox{-}on}}} \int_{0}^{T_{\mathrm{cal\mbox{-}on}}} y_n^{\mathrm{cal\mbox{-}on}}(t) \Bigl( y_1^{\mathrm{cal\mbox{-}on}}(t) \Bigr)^* dt \right\}.
\end{equation}
Beamforming weights are then set as
\begin{equation}
    w_n = e^{-j \hat{\Delta\phi}_n}.
\end{equation}
During observation, the combined array output is
\begin{equation}
    y_{\mathrm{combined}}^{\mathrm{PAC}}(t) = \sum_{n=1}^{N} w_n \, y_n^{\mathrm{obs}}(t).
\end{equation}

\subsubsection{Eigenvalue-Based Calibration (EVC)}
In EVC, the optimal beamforming weights are obtained by solving a generalized eigenvalue problem. The noise covariance matrix, estimated during the calibration-off period, is
\begin{equation}
    \mathbf{R}_n = \frac{1}{T_{\mathrm{cal\mbox{-}off}}} \int_{0}^{T_{\mathrm{cal\mbox{-}off}}} \mathbf{y}^{\mathrm{cal\mbox{-}off}}(t) \, \mathbf{y}^{\mathrm{cal\mbox{-}off}}(t)^\dagger dt,
\end{equation}
where $\mathbf{y}^{\mathrm{cal\mbox{-}off}}(t) = [y_1^{\mathrm{cal\mbox{-}off}}(t), \ldots, y_N^{\mathrm{cal\mbox{-}off}}(t)]^\top$. Similarly, the signal-plus-noise covariance during calibration-on is
\begin{equation}
    \mathbf{R}_{\mathrm{sn}} = \frac{1}{T_{\mathrm{cal\mbox{-}on}}} \int_{0}^{T_{\mathrm{cal\mbox{-}on}}} \mathbf{y}^{\mathrm{cal\mbox{-}on}}(t) \, \mathbf{y}^{\mathrm{cal\mbox{-}on}}(t)^\dagger dt.
\end{equation}
The signal covariance is then
\begin{equation}
    \mathbf{R}_s = \mathbf{R}_{\mathrm{sn}} - \mathbf{R}_n.
\end{equation}
The optimal weight vector $\mathbf{w}$ is obtained by solving
\begin{equation}
    \mathbf{R}_s \, \mathbf{w} = \lambda_{\max} \left( \mathbf{R}_n + \mathbf{I} \right) \mathbf{w},
\end{equation}
where $\mathbf{I}$ is the identity matrix and $\lambda_{\max}$ is the maximum eigenvalue. The corresponding eigenvector is normalized as
\begin{equation}
    \mathbf{w}_{\mathrm{norm}} = \frac{\mathbf{w}}{\left( \sum_{n=1}^{N} |w_n| \right)/N}.
\end{equation}
The combined array output during observation is then given by
\begin{equation}
    y_{\mathrm{combined}}^{\mathrm{EVC}}(t) = \mathbf{y}^{\mathrm{obs}}(t)^\dagger \, \mathbf{w}_{\mathrm{norm}}^*,
\end{equation}
with $\mathbf{y}^{\mathrm{obs}}(t) = [y_1^{\mathrm{obs}}(t), \ldots, y_N^{\mathrm{obs}}(t)]^\top$.

\textcolor{black}{The EVC method is inherently scalable through a hierarchical, multi-tier subarray architecture. Calibration begins at the local or neighborhood level, where small clusters of nearby IoT nodes (e.g., 100 devices) on their correlation matrices to estimate and align signal phases. Since each subarray operates independently, the matrix dimensions remain small, enabling real-time processing on local edge devices or gateways. Once local coherence is established, the outputs of these subarrays are aggregated recursively at higher levels—regional, national, and ultimately global—forming a layered calibration structure. This hierarchical approach ensures both scalability and high-fidelity phase alignment across the entire network.}

\subsection{SNR Calculation in the Observation Model}
During the observation period, after applying the calibration-derived beamforming weights to the downconverted signals, the combined output at the array is given by
\begin{equation}
    y_{\mathrm{combined}}^{\mathrm{obs}}(t) = \sum_{n=1}^{N} w_n\, y_n^{\mathrm{obs}}(t),
\end{equation}
where $w_n$ are the beamforming weights (obtained via PAC or EVC) and $y_n^{\mathrm{obs}}(t)$ is the complex baseband astronomical signal (including noise and any residual interference) at node $n$, as in Eq.~\eqref{eq:y_obs_base}.

To quantify the performance during observation, we define the signal power $P_s$ as the squared magnitude of the time-integrated coherent astronomical signal component. In practice, since the astronomical signal is much weaker than the noise, the signal power is obtained by isolating the known signal component (e.g., via a signal-only processing path):
\begin{equation}
    P_s = \left|\int_{0}^{T_{\mathrm{obs}}} y_{\mathrm{combined}}^{\mathrm{obs,sig}}(t)\, dt \right|^2,
\end{equation}
where $y_{\mathrm{combined}}^{\mathrm{obs,sig}}(t)$ denotes the beamformed signal when only the astronomical component is present. Similarly, the total received signal power during the observation period is defined as
\begin{equation}
    P_{\rm{obs}} = \left|\int_{0}^{T_{\mathrm{obs}}} y_{\mathrm{combined}}^{\mathrm{obs}}(t)\, dt \right|^2.
\end{equation}
The noise power, $P_n$, is then obtained by subtracting the signal power $P_s$ from the observed power:
\begin{equation}
    P_n = P_{\rm{obs}} - P_s.
\end{equation}
Consequently, the output signal-to-noise ratio (SNR) for the observation mode is expressed as
\begin{equation}
    \mathrm{SNR}_{\mathrm{obs}} = \frac{P_s}{P_n}.
\end{equation}

\subsection{Analysis Results}
\subsubsection{Comparison Between Uniform and Variable Nodes}
We simulate the performance of both calibration methods under two different scenarios. In the \textit{Uniform Node Characteristics} case, all nodes have identical antenna gains and noise figures (i.e., $\sigma_{G} = 0\,\text{dB}$ and $\sigma_{NF} = 0\,\text{dB}$), representing an ideal scenario with no hardware inconsistencies. In contrast, in the \textit{Variable Node Characteristics} scenario, nodes exhibit significant variability in antenna gains and noise figures, with $\sigma_{G} = 20\,\text{dB}$ and $\sigma_{NF} = 4\,\text{dB}$, simulating real-world conditions where hardware differences and antenna blockage effects are present. To specifically isolate the effects of node variability, we set the LO frequency offset $\Delta f$ and interference signals $i_{n}(t)$ to zero in both scenarios.

\begin{figure}[htp]
  \begin{center}
        \subfigure[]{
	\includegraphics[scale=\plotsize]{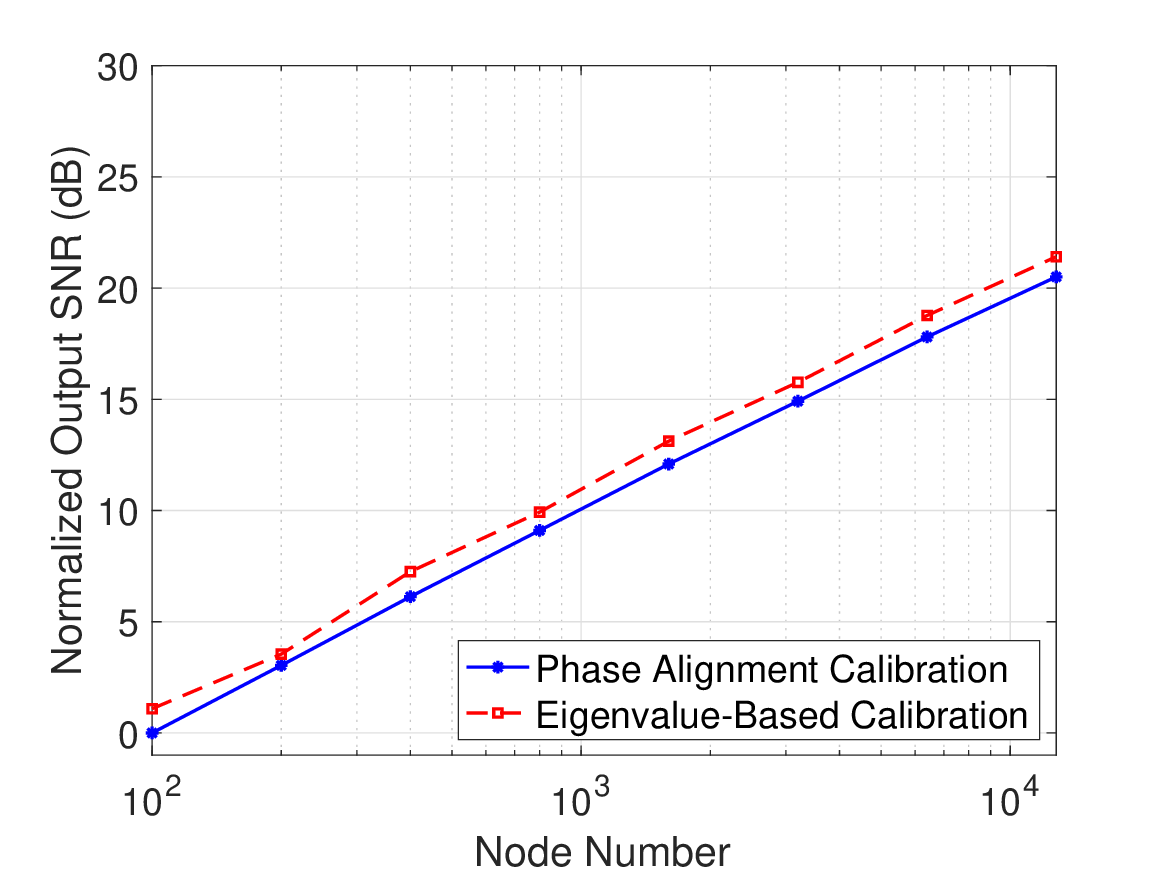}
	\label{fig:SNR_Comparison_0dB_0dB}}
        \subfigure[]{
	\includegraphics[scale=\plotsize]{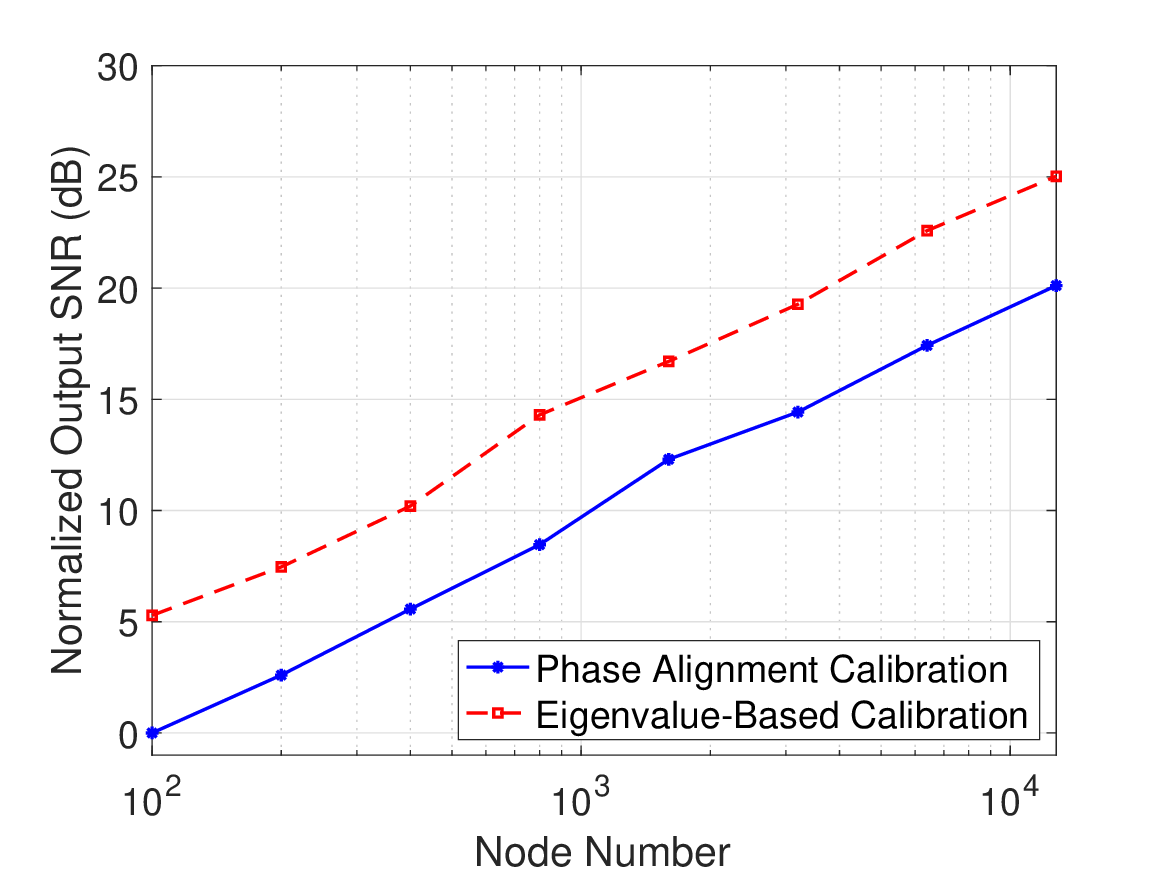}
	\label{fig:SNR_Comparison_20dB_4dB}}
  \end{center}
    \caption{Array output SNR versus the number of nodes for Phase Alignment Calibration (PAC) and Eigenvalue-Based Calibration (EVC) under (a) uniform node characteristics and (b) variable node characteristics.}
    \label{fig:SNR_comparison}
\end{figure}

For both scenarios, we vary the number of nodes $N$ from $100$ to $12,\!800$. The other analysis parameters are consistent across both scenarios: total array length $L = 1,\!000\,\text{km}$; center frequency $f_c = 1.42\,\text{GHz}$; bandwidth $B = 10\,\text{kHz}$; calibration signal power density $P_{\text{cal}} = 10^{-10}\,\text{W/m}^2$; astronomical signal power density $P_{\text{astro}} = 10^{-20}\,\text{W/m}^2$; mean antenna gain $\overline{G}_{\text{ant}} = -10\,\text{dB}$; mean noise figure $\overline{NF} = 4\,\text{dB}$; calibration time $T_{\text{cal}} = 1\,\text{ms}$ (with $T_{\text{cal-on}} = T_{\text{cal}} / 2$); observation time $T_{\text{obs}} = 1\,\text{ms}$; reference temperature $T_0 = 290\,\text{K}$.

The array output SNRs for both methods under the two scenarios are plotted in Fig.~\ref{fig:SNR_comparison}, where Fig.~\ref{fig:SNR_comparison}(a) corresponds to the uniform node characteristics and Fig.~\ref{fig:SNR_comparison}(b) corresponds to the variable node characteristics.

In the uniform node scenario shown in Fig.~\ref{fig:SNR_comparison}(a), both methods achieve similar array output SNRs across all values of $N$. This indicates that when the nodes are identical, the simpler PAC method is sufficient to achieve near-optimal performance.

In contrast, in the variable node scenario depicted in Fig.~\ref{fig:SNR_comparison}(b), the EVC method consistently outperforms the PAC method, demonstrating that EVC scales better with larger arrays under conditions of hardware variability.

Fig.~\ref{fig:SNR_diff_sigma_G} illustrates the array output SNR as a function of the standard deviation of the node gain $\sigma_{\rm{G}}$ when the number of node $N$ is 1,000. The EVC method consistently achieves higher output SNR than the PAC method. Moreover, the performance gap between the two methods widens as the gain variation increases, indicating that the EVC method is more robust to variations in node gain.

\textcolor{black}{In summary, while PAC is simpler and more computationally efficient, relying solely on phase-offset estimation, it does not account for amplitude variations or complex channel distortions. In scenarios with significant noise gain disparities across the array, EVC demonstrates superior performance by exploiting the full correlation matrix structure and isolating the dominant signal subspace.}

\begin{figure}[htp!]
      \centering
      \includegraphics[scale=\plotsize]{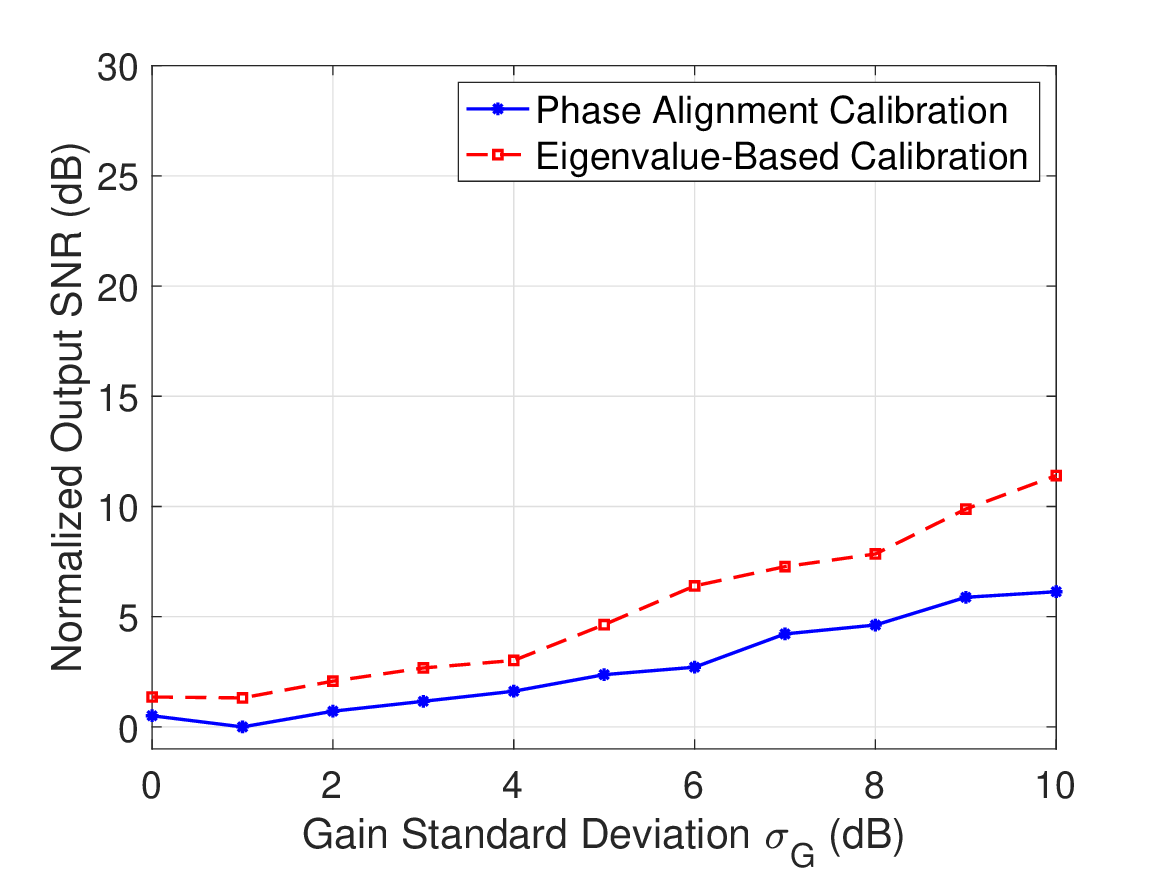}
      \caption{Comparison of array output SNR versus standard deviation of node gain $\sigma_{\rm{G}}$ for the Eigenvalue-Based Calibration EVC and Phase Alignment Calibration PAC methods.}
      \label{fig:SNR_diff_sigma_G}
\end{figure}

\subsubsection{Interference Cancellation}

A single-tone interference signal with power significantly higher than the thermal noise floor is assumed to impinge on the array from a direction different from the astronomical source. To mitigate the impact of such strong interference, SIC introduced in Section\ref{subsec:sic} is employed as a front-end processing step. Fig.~\ref{fig:Int_level} depicts the array output SNR as a function of the number of nodes for interference power densities $P^{\mathrm{int}}_{q}$ ranging from $10^{-4}\,\text{W/m}^2$ to $10^{-8}\,\text{W/m}^2$. The results show that, in observation mode, the output SNR varies only marginally over this wide range of interference levels. This behavior arises from the combined effects of SIC and distributed beamforming: while residual interference may persist at individual nodes after SIC, it is generally not phase-aligned across the array and therefore combines incoherently, whereas the desired astronomical signal combines coherently. These results suggest that, under the adopted modeling assumptions, the proposed SIC-based approach enables the IoT-enabled telescope to operate robustly in the presence of strong terrestrial communication interference, supporting effective spectrum sharing without significantly degrading astronomical observation performance.

\begin{figure}[htp!]
      \centering
      \includegraphics[scale=\plotsize]{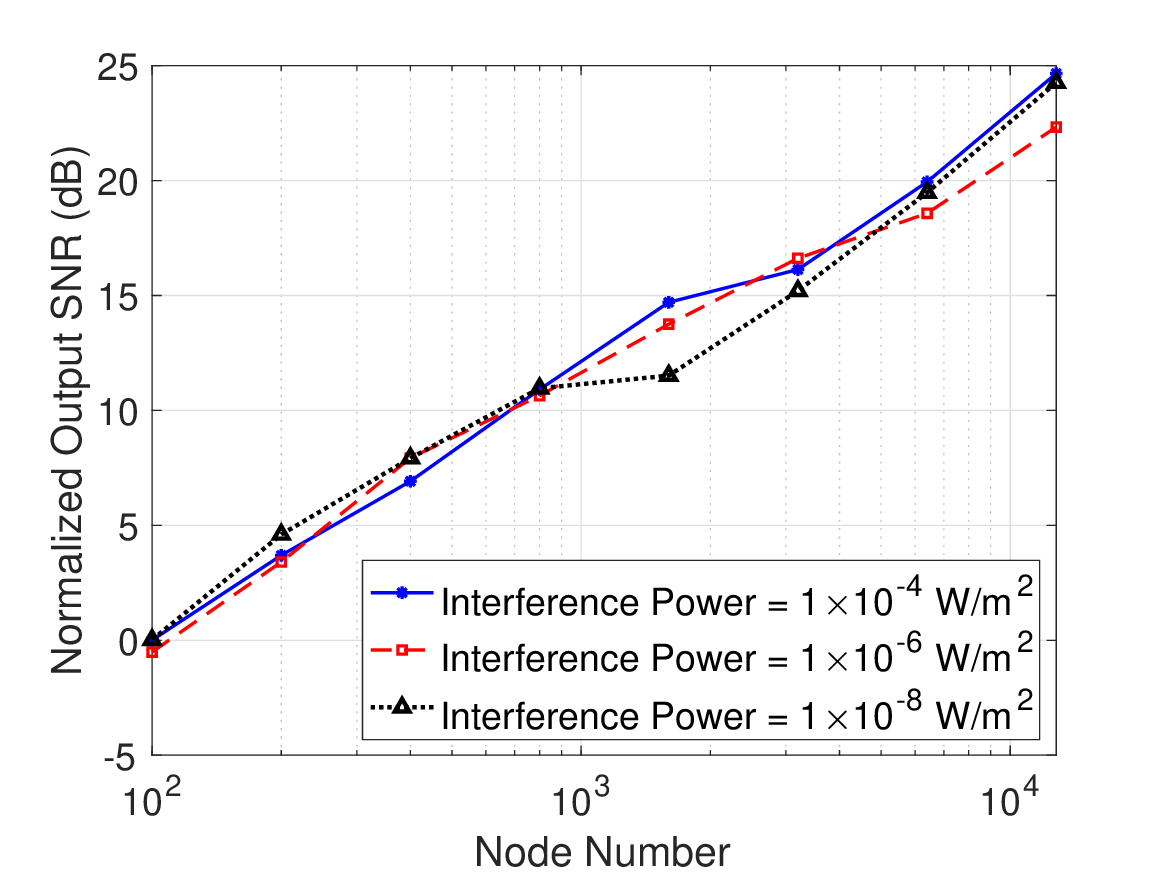}
      \caption{Comparison of array output SNR versus interference signal level.}
      \label{fig:Int_level}
\end{figure}

\subsubsection{Node LO Frequency Offset}  
\textcolor{black}{We evaluate the impact of local oscillator (LO) frequency offset $\Delta f$ on array coherence using a 50-node array. While fixing the calibration time $T_{\mathrm{cal}} = 1\,\mu\text{s}$, we vary the observation time $T_{\mathrm{obs}}$ from 50\,ms to 500\,ms and sweep $\Delta f$ from 0 to 2\,Hz. As shown in Fig.~\ref{fig:LO_offset}, when $T_{\mathrm{obs}} = 500\,\text{ms}$, a $\Delta f$ of 2\,Hz leads to a $\sim$7\,dB drop in output SNR. However, at $T_{\mathrm{obs}} = 50\,\text{ms}$, the SNR remains stable even at the same frequency offset. These results demonstrate that reducing the observation interval and increasing calibration frequency effectively mitigates LO drift, making the system compatible with IoT nodes disciplined by GPSDOs—typically accurate to $<$1\,Hz or $<$1\,ns~\cite{ettusUB210}. Moreover, over such short calibration–observation cycles, effects from atmospheric and ionospheric variations, as well as node motion, are negligible.}

\begin{figure}[htp!]
      \centering
      \includegraphics[scale=\plotsize]{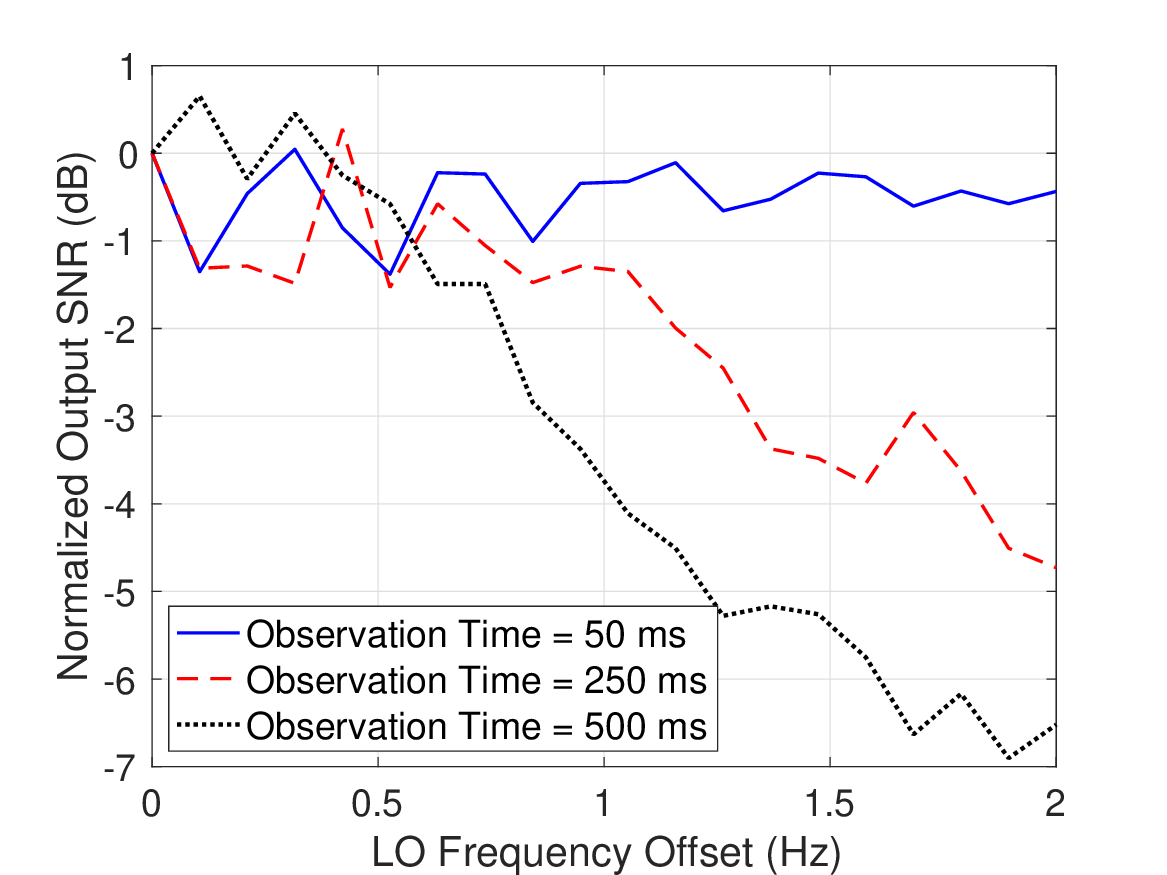}
      \caption{Comparison of array output SNR versus node LO frequency offset.}
      \label{fig:LO_offset}
\end{figure}

\section{Rough Performance Estimation and Comparative Analysis}

This section presents a comparative analysis of the proposed IoT-based radio telescope and the FAST, focusing on key performance metrics such as antenna gain, system noise temperature, sensitivity, field of view (FoV), bandwidth, beamwidth, and survey speed. The aim is to evaluate the potential of the IoT-enabled telescope concept relative to a state-of-the-art radio telescope.

\subsection{Performance Metrics}

The performance parameters considered for comparison include the antenna gain $G$, which represents the gain in the sky direction and incorporates the antenna radiation efficiency $\eta_{\text{rad}}$ and blockage effects $L_{\text{path}}$ due to environmental obstacles like buildings and trees. The system noise temperature $T_{\text{sys}}$ accounts for the total noise from both the environment and the receiver electronics. Sensitivity $S$ is defined as 
\begin{equation}
S = A / T_{\text{sys}}, 
\end{equation}
where $A$ is the antenna effective area. \textcolor{black}{The beamwidth or the angular resolution of the proposed IoT-telescope is approximated by 
\begin{equation}
\theta_{\text{bw,IoT}} \approx \frac{\lambda}{D_{\text{Earth}}},
\end{equation}
where $D_{\text{Earth}}$ is the diameter of the earth.}

The FoV is the angular area over which the telescope can detect signals, approximated by 
\begin{equation}
\text{FoV} = N_{\text{effb}} (\lambda / D)^2, 
\end{equation}
where $N_{\text{effb}}$ is the number of effective beams, $\lambda$ is the wavelength, and $D$ is the diameter of the telescope. Beamwidth $\theta_{\text{bw}}$ is given by $\theta_{\text{bw}} \approx \lambda / D$, representing the angular width of the main lobe of the antenna pattern. Bandwidth $B$ is the frequency range over which the telescope operates. Survey speed (SS) is calculated as 
\begin{equation}
\text{SS} = S^2  \text{FoV},
\end{equation}
reflecting the speed at which the telescope can survey the sky.

\subsection{FAST Parameters}
The FAST telescope is the world's largest single-dish telescope, featuring a diameter of 500 meters. It has a collection area of approximately 40,000 m$^{2}$, a system noise temperature of $T_{\text{sys,FAST}} = 20\,\text{K}$, and operates over a bandwidth of 800 MHz~\cite{nan2011five}.

Using these parameters, the antenna gain of FAST is calculated as 
\begin{equation}
G_{\text{FAST}} = \frac{4\pi A_{\text{FAST}}}{\lambda^2},
\end{equation}
where $A_{\text{FAST}}$ is the collecting area. The beamwidth or angular resolution is approximated by 
\begin{equation}
\theta_{\text{bw,FAST}} \approx \frac{\lambda}{D_{\text{FAST}}},
\end{equation}
and the field of view is given by 
\begin{equation}
\text{FoV}_{\text{FAST}} = N_{\text{effb}} \theta_{\text{bw,FAST}}^2,
\end{equation}
with $N_{\text{effb}} = 19$ for the cluster 19-element array feed system~\cite{nan2011five}.

\subsection{Proposed IoT-enabled telescope Parameters}

A rough estimation of the proposed IoT-based radio telescope's performance metrics can be derived from various front-end figures of merit. The antenna directivity $D_{\text{IoT}} $ for an individual IoT device is set at $0\,\text{dB}$, implying that these antennas are compact and designed to radiate power uniformly in all directions. The total antenna efficiency, encompassing radiation efficiency $\eta_{\text{rad}}$ and impedance matching, is approximately $-6\,\text{dB}$ \cite{bronckers2019benchmarking, ying2012antennas}, aligning with the efficiency typically observed in cellular phone antennas.

The average signal propagation loss $L_{\text{path}}$ due to building blockages is estimated at $-10\,\text{dB}$, reflecting the attenuation typically experienced from indoor to outdoor environments, such as GPS signal loss \cite{peterson1997measuring, micheli2014measurements, bui2020gps}. Given the omnidirectional radiation patterns of most single-antenna IoT devices, the external environmental noise temperature interfacing with the IoT receivers is expected to be around $300\,\text{K}$. The RF front-end noise temperature is primarily influenced by the first-stage LNA, which typically has a noise temperature of about $100\,\text{K}$, representative of commercial LNAs used in consumer electronics \cite{LNA_skyworks}. Since the receiver output SNR for most astronomical signals is below $0\,\text{dB}$, implementing a threshold near the receiver's noise floor can suppress interference from stronger communication signals. This strategic processing enables the network to treat interference from communication signals as background noise, which can subsequently be removed from the final data output.

The system's operational bandwidth is projected to be $4,000\,\text{MHz}$ based on the bandwidth of a commercial cell phone~\cite{appleStore}. With an estimated $N_{\text{IoT}} = 1 \times 10^{11}$ global IoT devices contributing to the beamforming process \cite{karunarathne2018wireless}. The individual device antenna gain at the sky direction is calculated as 
\begin{equation}
G_{\text{IoT}} = D_{\text{IoT}} L_{\text{path}} \eta_{\text{rad}}. 
\end{equation}
Multiplying by the number of devices, the cumulative antenna gain is 
\begin{equation}
G_{\text{IoT,total}} = G_{\text{IoT}} N_{\text{IoT}}/2.
\end{equation}

The beamwidth of the IoT-enabled telescope is approximated by $\theta_{\text{bw,IoT}} \approx \lambda / D_{\text{Earth}}$, where $D_{\text{Earth}} \approx 12,742\,\text{km}$ is the Earth's diameter. This results in an extremely narrow beamwidth due to the large effective aperture size. The field of view for the IoT-enabled telescope is considered to be $\text{FoV}_{\text{IoT}} = 4\pi$, covering the entire visible sphere of the earth.

\begin{table}[h]
\caption{Comparison of the Proposed IoT-enabled telescope with the FAST}
\label{tab:comparison}
\centering
\begin{tabular}{lcc}
\hline
\textbf{Parameter}                               & \textbf{FAST~\cite{nan2011five}}     & \textbf{IoT telescope} \\
\hline
Radiation Efficiency ($\eta_{\text{rad}}$)       & $0\,\text{dB}$                       & $-6\,\text{dB}$        \\
Building Blockage Loss ($L_{\text{path}}$)       & $0\,\text{dB}$                       & $-10\,\text{dB}$    \\
Antenna Gain ($G$) @1.4 GHz                       & $70.5\,\text{dB}$                    & $87\,\text{dB}$ \\
System Noise Temperature ($T_{\text{sys}}$)       & $20\,\text{K}$                       & $400\,\text{K}$ \\
Sensitivity ($S = A / T_{\text{sys}}$)           & $2,000\,\rm{m^{2}/K}$                & $5,550\,\rm{m^{2}/K}$ \\
Beamwidth ($\theta_{\text{bw}}$)                 & $8.5\times 10^{-4}\,\text{rad}$      & $1.7\times10^{-8}\,\text{rad}$ \\
Field of View (FoV)                              & $1.37 \times 10^{-5}\,\text{sr}$      & $4\pi\,\text{sr}$ \\
Bandwidth ($B$)                                  & $800\,\text{MHz}$                    & $4\,\text{GHz}$ (iPhone) \\
Survey Speed ($\rm{deg^{2} m^{4} K^{-2}}$)       & $1.8 \times 10^{5}$                  & $1.3 \times 10^{12}$ \\
Spectrum Sharing                                 & No                                   & Yes \\
Infrastructure Investments                       & High                                 & Low  \\
\hline
\end{tabular}
\end{table} 

\subsection{Performance Comparison}

Table~\ref{tab:comparison} summarizes the key performance metrics based on the above analysis. The center frequency is assumed to be 1.42 GHz. \textcolor{black}{It is emphasized that these metrics represent \emph{upper-bound estimates} obtained under the stated modeling assumptions, including idealized calibration, sufficient node number, and effective interference mitigation.}  Despite each individual IoT device having a low antenna gain due to the large antenna loss (-6 dB) and building blockage effect (-10 dB), the cumulative effect of aggregating signals from an immense number of devices results in a total antenna gain that is almost two orders of magnitude higher than that of the FAST. This significant increase is due to the vast number of participating IoT devices in the network. The system noise temperature of the IoT-enabled telescope is higher because IoT devices typically have greater noise figures, and the external environmental noise temperature is approximately equal to room temperature ($300\,\text{K}$). However, the IoT-enabled telescope offers a vastly larger field of view, effectively covering almost the entire sky simultaneously. In contrast, the FAST has a much smaller FoV due to its highly directional antenna. Consequently, the IoT-enabled telescope achieves a survey speed that is nearly seven orders of magnitude greater than that of the FAST, attributed to the large number of nodes and the expansive FoV. In practical terms, a survey that might take FAST approximate 10 million hours, or 1142 years, could be done in 1 hour with the proposed IoT-enabled telescope.

\begin{figure}[htp!]
      \centering
      \includegraphics[scale=\plotsize]{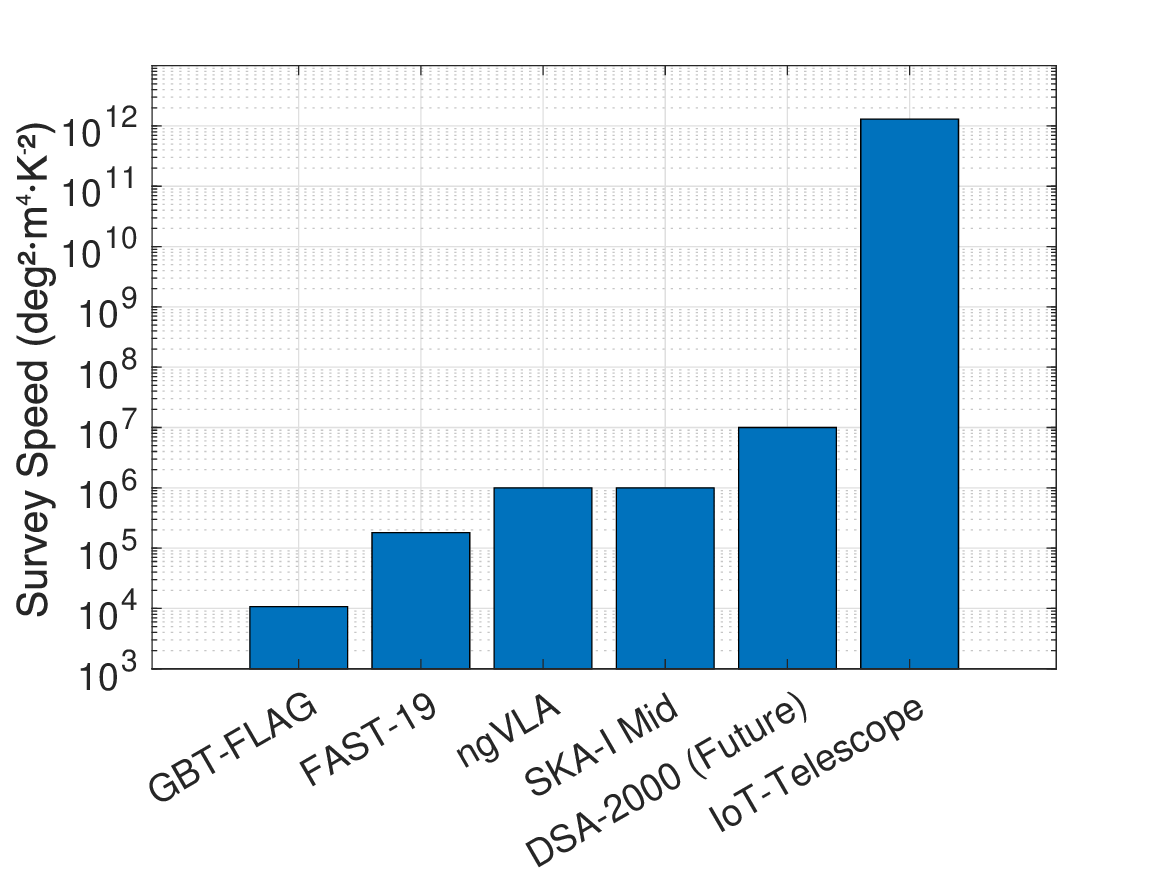}
      \caption{Comparison of approximate survey speed for several current and next‑generation major radio telescopes.}
      \label{fig:comparision_telescope}
\end{figure}

\textcolor{black}{Fig.~\ref{fig:comparision_telescope} compares the estimated survey speeds of several leading current and next-generation radio telescopes, including the Green Bank Telescope Focal L-Band Array (GBT-FLAG)~\cite{roshi2018performance}, the Five-hundred-meter Aperture Spherical Telescope 19-Beam Receiver (FAST-19)~\cite{li2016five}, the Next Generation Very Large Array (ngVLA)~\cite{selina2018next}, the Square Kilometre Array Phase 1 at Mid-Frequency (SKA-I Mid)~\cite{braun2019anticipated}, the Deep Synoptic Array 2000 (DSA-2000)~\cite{hallinan2019dsa}, and the proposed IoT-enabled telescope. Survey speed serves as the most important figure of merit for evaluating the capability of both single-dish and interferometric radio telescopes. Remarkably, the proposed IoT-enabled telescope is projected to achieve a survey speed nearly five orders of magnitude greater than that of the DSA-2000, which is currently under development and expectes to have the highest survey speed.}

Furthermore, it is important to note that conventional radio telescopes—both current and future—require substantial infrastructure investments, are engineered for low-noise performance, and must be situated in radio quiet zones. In contrast, IoT-enabled radio telescopes leverage existing IoT infrastructure as receivers and the potential use of existing low-orbit satellites as calibrators, thereby eliminating the need for expensive, dedicated systems. Consequently, these systems can operate in interference-rich environments, enabling efficient spectrum sharing with wireless communication networks.

\section{Conclusion}
This paper introduces an innovative approach to radio astronomy by leveraging the global IoT network to build a distributed radio telescope. Utilizing a vast array of dispersed nodes, the system overcomes the spectrum sharing and coverage constraints of traditional single-aperture telescopes. Digital beamforming, coupled with GPS-synchronized calibration via satellite beacons, effectively aligns astronomical signals and mitigates phase errors. Performance assessments indicate that Eigenvalue-Based Calibration (EVC) outperforms Phase Alignment Calibration (PAC) under variable node conditions, while successive interference mitigation method minimizes interference. Furthermore, the adverse impact of local oscillator (LO) frequency offsets is reduced by shortening observation durations. In comparison to the 500-meter Aperture Spherical Telescope (FAST), the IoT-enabled design achieves over two orders of magnitude higher antenna gain and nearly seven orders of magnitude faster survey speed, thanks to its extensive effective aperture and all-sky field-of-view. These findings validate a cost-effective, scalable platform for radio astronomy, with future work focusing on prototyping with software-defined radios and addressing challenges in synchronization, data processing, and interference management.

\bibliographystyle{IEEEtran}
\bibliography{IoT_telescope.bib}


\end{document}